\documentclass[%
 aip,
rsi,%
 amsmath,amssymb,
 reprint,%
 floatfix,
]{revtex4-1}
\usepackage{graphicx}
\usepackage{dcolumn}
\usepackage{bm}

\begin{document}

\preprint{AIP/123-QED}

\title[Cryogenic single-port calibration for superconducting microwave resonator measurements]{Cryogenic single-port calibration for superconducting microwave resonator measurements}
\author{Haozhi Wang}
\email{haozhi.wang@colorado.edu}
\affiliation{
Department of Physics, University of Colorado, Boulder, CO 80309, USA
}
\affiliation{ 
National Institute of Standards and Technology, Boulder, CO 80305, USA
}%
\affiliation{ 
JILA, University of Colorado, Boulder, CO 80309, USA
}%
\affiliation{ 
Boulder Cryogenic Quantum Testbed, University of Colorado, Boulder, CO 80309, USA
}
 
\author{S. Singh}
\email{suren\_singh@keysight.com}
\affiliation{ 
Keysight Technologies,
Santa Rosa, CA 95403, USA 
}%

\author{C.R.H. McRae}
\affiliation{
Department of Physics, University of Colorado, Boulder, CO 80309, USA
}
\affiliation{ 
National Institute of Standards and Technology, Boulder, CO 80305, USA
}%
\affiliation{ 
JILA, University of Colorado, Boulder, CO 80309, USA
}%
\affiliation{ 
Boulder Cryogenic Quantum Testbed, University of Colorado, Boulder, CO 80309, USA
}

\author{J.C. Bardin}
\affiliation{
Google Quantum AI,
Mountain View, CA 94043, USA
}
\affiliation{ 
Department of Electrical and Computer Engineering, University of Massachusetts Amherst, Amherst, MA 01003-9292, USA
}

\author{S.-X. Lin}
\affiliation{
Department of Physics, University of Colorado, Boulder, CO 80309, USA
}
\affiliation{ 
National Institute of Standards and Technology, Boulder, CO 80305, USA
}
\affiliation{ 
JILA, University of Colorado, Boulder, CO 80309, USA
}
\affiliation{ 
Boulder Cryogenic Quantum Testbed, University of Colorado, Boulder, CO 80309, USA
}

\author{N. Messaoudi}
\affiliation{ 
Keysight Technologies, Santa Rosa, CA 95403, USA 
}%
\affiliation{ 
Institute for Quantum Computing and Electrical and Computer Engineering, University of Waterloo, Ontario, Canada
}%

\author{A.R. Castelli}
\affiliation{ 
Lawrence Livermore National Laboratory, Livermore, CA 94550, USA
}

\author{Y.J. Rosen}
\affiliation{ 
Lawrence Livermore National Laboratory, Livermore, CA 94550, USA
}

\author{E.T. Holland}
\altaffiliation{Present address: Quantum R$\&$D Center, Keysight Technologies Inc., 
Cambridge, Massachusetts 02142, USA
}
\affiliation{ 
Lawrence Livermore National Laboratory, Livermore, CA 94550, USA
}

\author{D.P. Pappas}
\affiliation{ 
National Institute of Standards and Technology, Boulder, CO 80305, USA
}%
\affiliation{
Department of Physics, University of Colorado, Boulder, CO 80309, USA
}

\author{J.Y. Mutus}
\affiliation{
Google Quantum AI,
Mountain View, CA 94043, USA
}
\affiliation{ 
Boulder Cryogenic Quantum Testbed, University of Colorado, Boulder, CO 80309, USA
}

\date{\today}

\begin{abstract}
Superconducting circuit testing and materials loss characterization requires robust and reliable methods for the extraction of internal and coupling quality factors of microwave resonators. A common method, imposed by limitations on the device design or experimental configuration, is the single-port reflection geometry, i.e. reflection-mode. However, impedance mismatches in cryogenic systems must be accounted for through  calibration of the measurement chain while it is at low temperatures. In this paper, we demonstrate a data-based, single-port calibration using commercial microwave standards and a vector network analyzer (VNA) with samples at millikelvin temperature in a dilution refrigerator, making this method useful for measurements of quantum phenomena. Finally, we cross reference our data-based, single-port calibration and reflection measurement with over-coupled 2D- and 3D-resonators against well established two-port techniques corroborating the validity of our method.
%
\end{abstract}

\maketitle

\section{Introduction}

Microwave resonators are a key element in superconducting quantum computing. These structures are widely used as filters,~\citep{sete2015quantum,reed2010fast,jeffrey2014fast,bronn2015reducing} readout elements,~\citep{Blais2004,Wallraff2004} and amplifiers,~\citep{castellanos2007widely,bergeal2010phase,mutus2014strong,roy2015broadband,bal2020overlap} amongst other uses. In addition, the measurement of the resonator internal quality factor, $Q_i$, can be used to evaluate circuit performance and characterize both materials loss~\citep{Martinis2005,GaoThesis,mcrae2020materials} and experimentally induced loss.~\citep{Song2009,Chiaro2016,Sage2011} The frequency response of frequency-multiplexed, two-dimensional (2D) resonators is typically measured in the hanger geometry with a transmission line (see Fig.~\ref{fig:circuit_diagram} below), referred to as hanger-mode herein.\citep{Gao2008b,McRae2020} In the hanger-mode, multiple resonators can be coupled to a single feedline and data fitting is simplified due to the background self-leveling. Therefore, techniques can be implemented to extract $Q_i$ accurately from the embedding circuit.~\citep{Khalil2012,Megrant2012,guan2020network,Probst2015}  

On the other hand, it is often desired to measure devices using a reflection-mode. This is typically implemented for 2D readout/drive-resonators for cross-resonance gates,~\citep{chow2014implementing,ware2019cross} Josephson parametric amplifiers,~\citep{manuelthesis} and  three-dimensional (3D) resonant cavities.~\citep{kudra2020high,chakram2020seamless} The main advantage of the reflection geometry is that all of the intracavity signal power is available for measurement. This gives nominally a factor of 2 improvement of the signal to noise ratio (SNR) relative to the hanger-mode, where signal can leak back to the feedline. From a practical perspective, the reflection geometry is nice because it requires only a single connection to the device under test (DUT). Thus, multi-sample measurement is simplified because a single-pole switch can be utilized to test multiple devices.  The problem addressed herein is that there is currently not a method to account for impedance mismatches in the reflection-mode measurement chain.

Previously, there have been demonstrations of one- and two-port calibrations for low temperature measurements, but they have several limitations: some rely on simulation results rather than actual measurements and target specific applications;~\citep{jun2004microwave} and others rely on room temperature calibration for DUTs that operate at cryogenic temperatures. An additional possible variable for this latter set of calibrations is the actual operating temperature of the DUTs, i.e. L-He ($\approx$ 4 K)  vs. base temperature of a DR ($T_{base}\approx10$~mK).~\citep{slichter2009millikelvin} Previous work by Ranzani~\emph{et al.}, 2013 demonstrated a two-port, through-reflect-line (TRL) calibration method~\citep{ranzani2013two} that can be applied to DUTs operated at $T_{base}$.~\citep{yeh2013situ} However, this technique relies on a relatively complicated scheme with two multi-position switches and four separate $S_{21}$ measurements to acquire the full $2\times2~S$ parameters required for calibration, as well as additional external data processing software.

In this paper we demonstrate a single-port, short-open-load (SOL) calibration at $T_{base}$ that requires only one switch and no external software. To accomplish this, we change setup of the VNA configuration so that the reflected signal is measured by the same port that sends the output signal. In this way, only a single port of the VNA is required and the built-in calibration functions can be used directly for post-processing the measurement data. Importantly, we note that this configuration can be extended to two-port applications and beyond. 

This paper is structured as follows: In Sec.~\ref{sec:resonantor_model}, we briefly revisit hanger and reflection-mode measurements and use simulation results to demonstrate the impact of calibration on $Q_i$ extraction in the reflection-mode measurement; In Sec.~\ref{sec:cal_theory}, we describe the theory for single-port calibration using a VNA and calibration standards; In Sec.~\ref{sec:experiment_Setup}, we illustrate the experimental setup and process for single-port data-based  calibration and verify the accuracy of the calibration by measuring different test standards; 
In Sec.~\ref{sec:Cal_Applied_to_Reflection}, we experimentally demonstrate the improvement on $Q_i$ measurement accuracy by applying the calibration to 2D and 3D superconducting microwave resonators; and lastly, in Sec.~\ref{sec:conclusion}, we summarize our work and give the outlook for future developments. 

\section{The Resonator Model}
\label{sec:resonantor_model}

Resonators measured in the hanger- and reflection-modes both exhibit a Lorentzian-shaped response at resonance. However, various parameters in the fitting, e.g. background and phase of response, in the two modes due to impedance mismatches and non-idealities in the microwave embedding circuit are very different. This section outlines the basic parameters required to fit the response from these two modes.

\subsection{Hanger-mode measurement}
\label{sec:hanger_mode_measurement}

\begin{figure}[htb!]
\includegraphics[width=\linewidth]{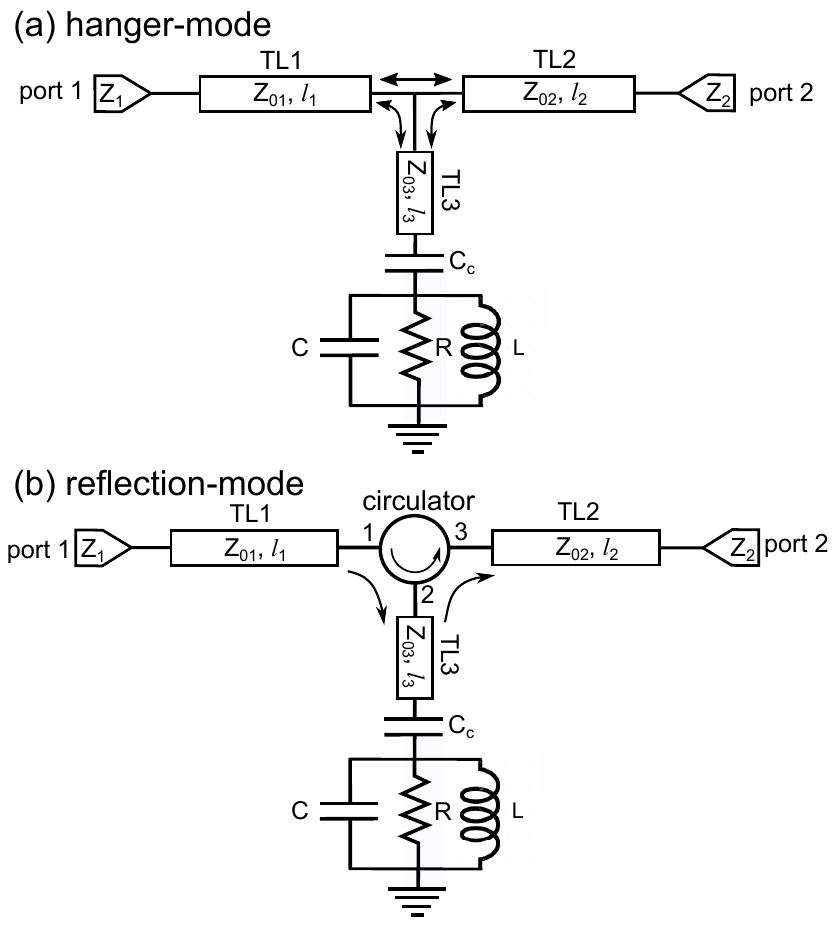}
\caption{\label{fig:circuit_diagram} Circuit diagram of (a) hanger-mode and (b) reflection-mode measurement of a parallel LCR resonator capacitively coupled to the environment with arrows indicating the direction of signal travel. Transmission lines are labeled as TL$x$, where $x$ is the number index, with characteristic impedance $Z_{0x}$ and electrical length $l_{x}$. Port impedance is labeled as $Z_x$. For hanger-mode, $l_3$ is often set to be $0$ for 2D planar resonator (which is the case in this paper), or approximately $13~$mm when using an sub-miniature version A (SMA) Tee for 3D cavity (see Appendix~\ref{sec:tee}).
}
\end{figure}

The hanger-mode is realized by coupling the resonators to a microwave feedline through a capacitor or an inductor. When measured in a two-port setup [Fig.~\ref{fig:circuit_diagram}(a)], the transmission $S_{21}(\omega)$ can be represented by 
\begin{equation}
\label{eq:S21 mismatch}
    S_{21}(\omega)=1-\frac{\frac{Q}{Q_c} e^{i\theta}}{1+2 i Q \frac{\delta \omega}{\omega}}
\end{equation}
near resonance frequency $\omega_0$, where $Q$ is the total quality factor, $Q_c$ is the coupling quality factor, $\delta \omega=\omega-\omega_0$ and  $\theta$ is a rotation angle with respect to the off-resonance point caused by imperfect coupling. As shown in Eq.~\ref{eq:S21 mismatch}, $S_{21}(\omega)$ is essentially just a Lorentzian function subtracted from unity. Its trajectory on the complex plane is a circle with the off-resonance point near $(1,0)$ in the complex plane. Here, the cable delay, loss, attenuation, and amplification in the measurement chain are removed by normalizing to unity. Hence, the diameter of the circle is $Q/Q_c \leq 1$ and $Q_i$ of the resonator can be found through
\begin{equation}
\label{eq:Qi original}
    \frac{1}{Q_i}=\frac{1}{Q}-\frac{1}{Q_c}.
\end{equation} 
Equation~\ref{eq:S21 mismatch} is often referred to as an ``ideal resonance".~\cite{Probst2015} However, its magnitude will pass unity if $\theta \neq 0$, which appears to violate energy conservation. As a result, Eq.~\ref{eq:S21 mismatch} returns an incorrect, inflated value of $Q_i$.
This is due to the fact that the normalization process increased the diameter of the circle by $1/\cos\theta$. To account for this, we implement the diameter correction method (DCM), where $Q_i$ is corrected as~\citep{Khalil2012} 
\begin{equation}
\label{eq:Qi correct}
    \frac{1}{Q_i}=\frac{1}{Q}-\frac{\cos\theta}{Q_c}.
\end{equation}
In Appendix~\ref{sec:resonator diameter correction}, we briefly discuss the effect of DCM by reviewing the resonance circle and the limitations on the rotation angle $\theta$. 

\subsection{Reflection-mode measurement}\label{subsec:reflection measurement}

The reflection-mode is typically realized by directly measuring the reflected signal from a resonator shunted to ground. Unlike the hanger-mode, the reflection-mode is technically a single-port, $S_{11}$, measurement. However, when measuring the quantum-level response of samples at $T_{base}$ in a dilution refrigerator, the input line is normally heavily attenuated in order for thermalization and noise reduction. Therefore, the reflected signal is very low, and would be further attenuated if an attempt was made to measure it coming back up the input line. Hence, directional devices such as circulators or directional-couplers are used to route the reflected signal back out of the DR along an amplified output line. In this manner, the measurement is converted to a two-port $S_{21}$ measurement [Fig.~\ref{fig:circuit_diagram}(b)]. Near resonance, $S_{21}$ can be expressed by~\citep{kudra2020high}
\begin{equation}
\label{eq:S21 reflection_rotation}
    S_{21}(\omega)=1-\frac{2\frac{Q}{Q_c} e^{i\theta}}{1+2 i Q \frac{\delta \omega}{\omega}}.
\end{equation}
Eq.~\ref{eq:S21 reflection_rotation} differs from Eq.~\ref{eq:S21 mismatch} by a factor of 2 in the Lorentzian term. Unfortunately, despite the similarities in the two forms, it is not possible to normalize the reflection-mode signal so a DCM-method is not applicable.
\begin{figure}
\includegraphics[width=\linewidth]{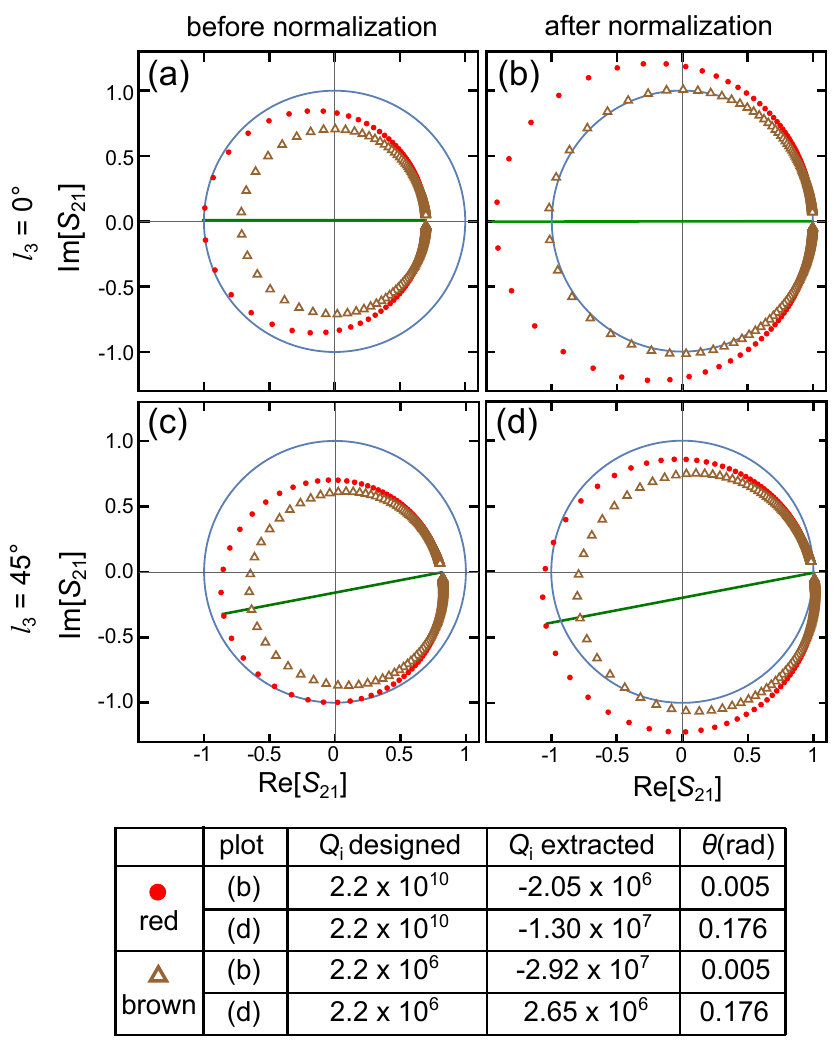}
\caption{\label{fig:Reflection_Plot} (Color online) Simulated $S_{21}$ of a parallel LCR resonator measured in reflection-mode. Electrical length varies between $l_3=0^{\circ}$ [(a), (b)] and $45^{\circ}$ [(c), (d)] at $6~$GHz. Circuit model is shown in Fig.~\ref{fig:circuit_diagram}(b). $S_{21}$ is plotted on the complex plane before [(a), (c)] and after [(b), (d)] normalizing the off-resonance point to 1. Simulation parameters: $Z_1=20~\Omega$, $Z_2=120~\Omega$, $Z_{01,02,03}=50~\Omega$, $l_{1,2}=90^{\circ}$ at $6~$GHz, $l_3=0^{\circ}$ for (a) and (b) and $45^{\circ}$ for (c) and (d) at $6~$GHz, circulator port isolation $300~$dB with no insertion loss, $C=580~$fF and $L=1.2~$nH sets the resonance frequency to be near $6~$GHz, $R=10^{12}~\Omega$ for designed $Q_i$ of $2.2\times10^{10}$ 
and $10^8~\Omega$ for designed $Q_i$ of $2.2\times10^{6}$.
$C_c=1~$fF sets $Q_c$ to be approximately $3\times10^5$. The blue circle is the unity circle. The green line is the diameter from off-resonance point to the resonance point for the red circle as a guide to the eye for the rotation angle. Eqs.~\ref{eq:Qi correct} and~\ref{eq:S21 reflection_rotation} are used for extraction of $Q_i$ shown in table. 
} 
\end{figure}

In Fig.~\ref{fig:Reflection_Plot}, we show the simulated results for a parallel LCR resonator with fixed $Q_c$ and $\omega_0$ but varied $Q_i$ in two different impedance mismatch scenarios: the input port and output port impedance are set to $20~\Omega$ and $120~\Omega$ (these extreme values are chosen to better demonstrate the effect caused by mismatches) for both cases but the electrical length of TL3 is set to $0^\circ$ for first scenario and $45^\circ$ (at $6~$GHz) for second scenario. 
The mismatch suppresses the transmission background so the off-resonance point before normalization is lower than unity. After normalization, the diameters are magnified and the resonance circles can pass unity. This results, again, in an incorrect, inflated value of $Q_i$. However, in the first scenario [(a)-(b)], there is almost no rotation in the signal and in the second scenario [(c)-(d)], the reported rotation angle is too small, showing that the DCM is not useful. The result is, due to the impedance mismatch, the extracted $Q_i$ are always higher than the actual value. In fact, in the fit for the over-coupled resonator (red dotted circle) can even return negative values of $Q_i$, thus indicating a false, unphysical gain. Experimental results in Sec.~\ref{subsec:reflection measurement} also demonstrate this effect.

Compounded with the effects of impedance-mismatches, the imperfect isolation of the circulator used for signal separation, normally $\approx 20~$dB, also affects the measurement accuracy. In Table~\ref{tab:table1}, we show $Q_i$'s extracted from simulations of various isolation values. Port impedance, transmission line characteristic impedance and circulator port impedance are all set to $50~\Omega$. It can be seen that the extracted $Q_i$ deviates significantly from the actual value as isolation decreases from near infinity to $20~$dB. That means the leakage from port 1 of circulator to port 3 will interfere with signal coming from port 2 to port 3. The error caused by the non-ideal isolation is similar to that caused by the directivity in coupler-based signal separation used in S-parameter measurements.\citep{app-note2} 
\begin{table}
\caption{\label{tab:table1}Error induced by imperfect circulator. Here, we set $Z_{1,2}=Z_{01,02,03}=50~\Omega$, $l_{1,2}=90^{\circ}$ at $6~$GHz, $l_3=0^{\circ}$, $C_c=1~$fF, $C=580~$fF, $L=1.2~$nH, $R=10^8~\Omega$.
}
\begin{ruledtabular}
\begin{tabular}{cccc}
Isolation & $Q_i$ designed & $Q_i$ extracted & $\theta$\\
\hline
300 dB & $2.2\times10^6$  & $2.2\times10^6$ & 0\\
30 dB &$2.2\times10^6$ & $1.73\times10^6$& 0  \\
23 dB & $2.2\times10^6$ &$1.39\times10^6$& 0  \\
20 dB & $2.2\times10^6$ &$1.22\times10^6$& 0  
\end{tabular}
\end{ruledtabular}
\end{table}

Due to the difficulty of accounting for impedance mismatches and imperfect circulator isolation, single-port calibration at cryogenic temperatures that can move the reference plane to the sample input port is critical for the accuracy of reflection measurements. Below, we show a method of calibrating out these effects by reconfiguring the front panel connections, i.e. test set connections, and using the built-in functions of a commercial VNA. 

\section{Single Port Measurement and Calibration Using a VNA}
\label{sec:cal_theory}
\begin{figure}[ht]
   \centering
   \includegraphics[width=0.48\textwidth]{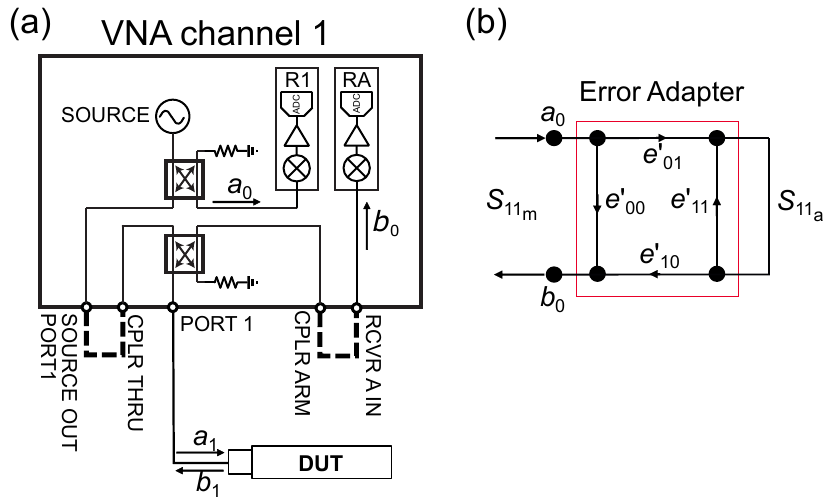}
   \caption{(a) Diagram of single-port measurement using a VNA. $a_0$ is the applied signal from the source, $b_0$ is the measured reflected signal, $a_1$ is the applied signal on the DUT, and $b_1$ is the reflected signal from the DUT. The dashed lines represent front panel jumper wires that can be removed to reconfigure the VNA. Some commercial VNAs do not have front jumpers, but the source, signal separation devices and receivers are universal. (b) single-port measurement error model. The red-outlined box represents the error adapter (also known as error box) from the measurement port to the DUT. $S_{11_a}$ is the actual $S_{11}$ and $S_{11_m}$ is the measured $S_{11}$. 
   }
    \label{fig:VNA_block}
\end{figure}

In order to understand the calibration protocol necessary for integrating the directional nature of microwave measurements in a DR it is necessary to have a more detailed understanding of the internal workings of a VNA. Here, we briefly introduce the VNA setup, sources of error, and calibration for a single port measurement. Fig.~\ref{fig:VNA_block}(a) shows the block diagram of one channel of a typical VNA. A source creates an outgoing wave that is split by a  directional coupler into two signal paths. One goes to receiver R1 and is measured as $a_0$. The other is routed to the SOURCE OUT port, then passes a jumper cable to the CPLR THRU port. It then travels from VNA PORT 1 to the DUT's input port and is labelled as $a_1$. The reflected signal from the DUT, $b_1$, travels through another directional coupler to the CPLR ARM port, then through a jumper cable to the RCVA A IN port, and then back to receiver RA to be measured as $b_0$. The ratio $b_0/a_0$ is the measured $S_{11_m}$ while $b_1/a_1$ represents the actual $S_{11_a}$ of the DUT. 

The measurement accuracy of $S_{11}$  using the VNA is dependent on three key performance factors: directivity of the system ($e'_{00}$), i.e. the ability of the VNA to separate waves $a_1$ and $b_1$; reflection tracking ($e'_{01}  e'_{10}$), i.e. the ability of the system to track the frequency response of the reflected signal; and source matching ($e'_{11}$), i.e. the ability of the system impedance to match that of the DUT. Because these quantities are unique to the system architecture, they are considered to be systematic errors that can be quantified and accounted for by using a set of known calibration standards.~\citep{rytting1980analysis}
Fig.~\ref{fig:VNA_block}(b) shows the flow graph model~\citep{garelli2012unified} used to describe these systematic quantities and their relationship to the measured quantities of a single-port device.

We start by relating the measured $S_{11_m}$ to the actual $S_{11_a}$ using
\begin{equation} \label{1_port}
S_{11_m} =\frac{b_0}{a_0} = e'_{00} +\frac {\left(e'_{01}  e'_{10}\right)S_{11_a}}{1-e'_{11} S_{11_a}}.
\end{equation} 
The error terms can be extracted and calibrated by independently measuring three known devices. These are typically referred to as calibration standards. Commonly, a short, open and load are used. The standards are pre-measured with high accuracy. There are two common methods used to describe the standards: polynomial calibration that use a set of closed-form polynomial equations to describe the standards' physical properties,~\citep{app-note1,bianco1978evaluation,blackham2005latest} and data-based calibration that uses the measurement data directly.~\citep{app-note1,bianco1978evaluation,blackham2005latest} In this paper, we use the data-based calibration method.

\section{Experimental Setup}
\label{sec:experiment_Setup}

The experimental setup for calibration and resonator measurements is shown in Fig.~\ref{fig:fridge_wiring}. The VNA test set is reconfigured by removing the two jumper cables [see Fig.~\ref{fig:VNA_block}(a) and Sec.~\ref{sec:calibration_process}] and connecting the SOURCE OUT and RCVR A IN ports to the DR input and output respectively. The reflection coupler is being bypassed here and replaced with a circulator down at $10~$mK. By doing this, the loss prior to the DUT can be moved to the other side of the sampler, greatly improving sensitivity of measurements and calibration. A DR with base temperature below $15~\mathrm{mK}$ is used. On the input side, the setup includes $20~\mathrm{dB}$ attenuation on the $3~\mathrm{K}$ plate and another $40~\mathrm{dB}$ attenuation on the mixing chamber followed by a low-pass filter with a linear filtering slope ($1~\mathrm{dB}/\mathrm{GHz}$). The total combined cable loss on the input side is roughly $10~$dB. Because the input line is heavily attenuated, a circulator, with a working frequency band of $4-12~$GHz, is used to route the input signal to a six-way mechanical switch. Six nominally identical cables are used to connect the switch channels to different DUTs. The output signal is routed by the circulator to a double stage isolator, followed by a $40~\mathrm{dB}$-amplification 
high-electron-mobility transistor (HEMT) with a $4~\mathrm{GHz}$ to $8~\mathrm{GHz}$ bandwidth, and then another $30~\mathrm{dB}$ amplification at the DR output. In this setup, RA receives the output signal while R1 still calibrates the source. Effectively, the VNA measures the reflection using a single-port $S_{11}$ measurement instead of the two-port $S_{21}$ measurement described in Sec.~\ref{subsec:reflection measurement}. This method allows for a relatively simple calibration using typical built-in functionality of VNAs that calculates the error terms without the need of external software.
\begin{figure}
\label{fig:fridge_wiring}
\includegraphics[width=\linewidth]{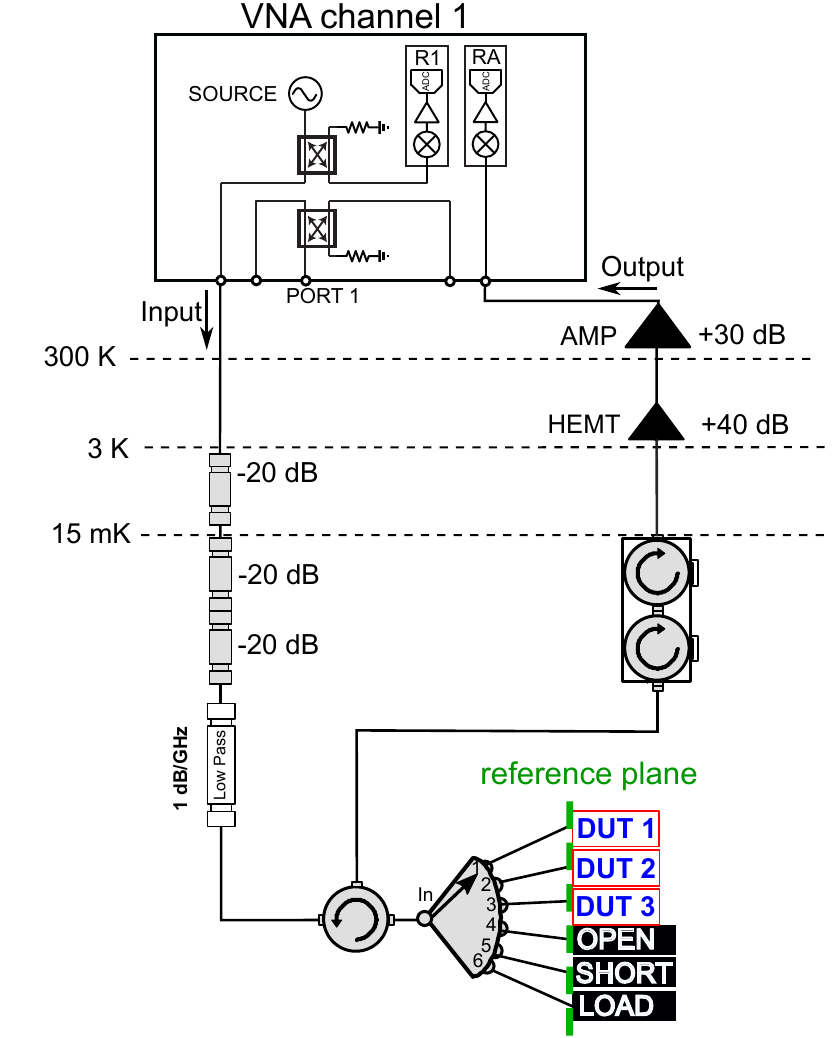}
\caption{\label{fig:fridge_wiring} (Colored online) DR wiring diagram for cryogenic microwave measurement. The VNA is reconfigured with corresponding front jumper wires (shown in Fig.~\ref{fig:VNA_block}) removed for reflection measurement. Temperature plates with no devices attached are excluded. The reference plane is marked by the green dashed line. Cryogenic switch channels 1-3 are connected to testing and verification DUTs. Open, short and load standards on channels 4-6 are calibration standards and are left in place for all cool-downs in this work.
}
\end{figure}

\subsection{Calibration Protocol}\label{sec:calibration_process}

A set of cryogenically compatible open-, short- and load-standards were initially measured at ambient temperature to create a data-based calibration kit. The standards are then connected to the switch at $T_{base}$ of the DR, as shown in Fig.~\ref{fig:fridge_wiring}. The only assumption, i.e. that each port of the switch and the cables connecting the devices are identical, was validated by switch repeatability tests and ambient temperature channel comparisons (see Appendix~\ref{sec:switch} and References ~\citep{ranzani2013two,yeh2013situ}). 
A $1~\mathrm{kHz}$ intermediate frequency (IF) bandwidth was used, with 2,001 data points, to achieve desired resolution and dynamic range and to limit noise. We reset/set the switch position by sending short-duration ($15~\mathrm{ms}$) DC pulses to disconnect/connect each channel, respectively. The switch operation voltage at $T_{base}$ is $4~\mathrm{V}$. These two successive operations tend to heat the mixing chamber by approximately $1~\mathrm{mK}$, and a wait time of approximately 25 to 30 minutes is required to allow the DR mixing chamber to reach a stable temperature. 

\subsection{Calibration Results and Verification}
\label{sec:verfication}

\begin{figure}
   \centering
   \includegraphics[width=\linewidth]{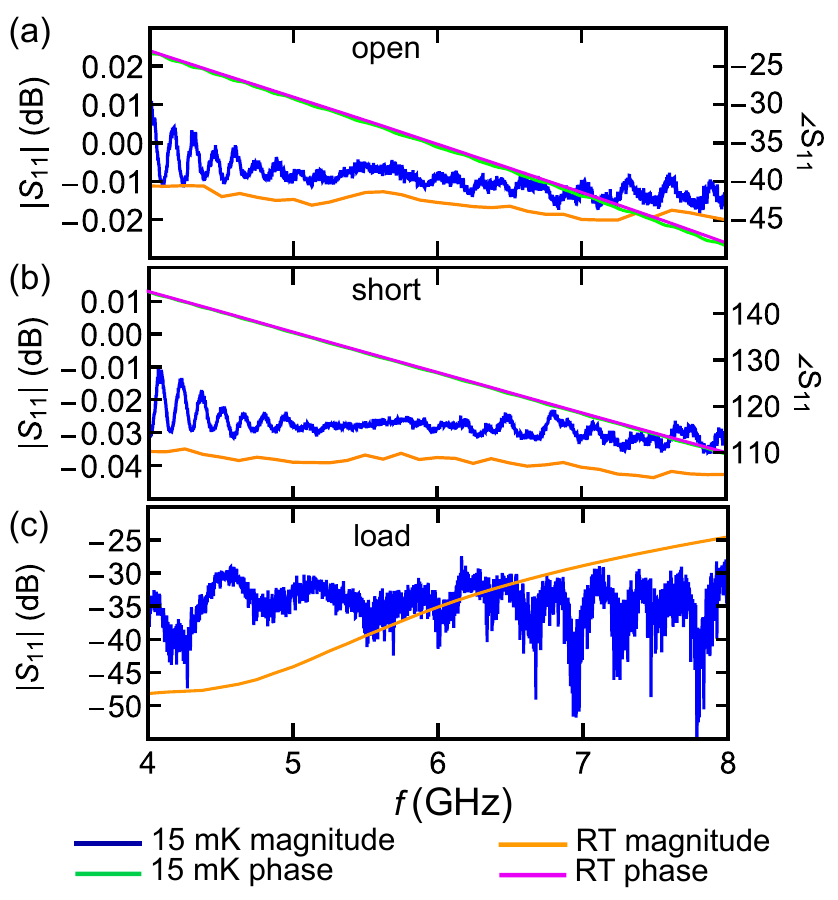}
   \caption{(Colored Online) Measurements of three calibration standards at base temperature, open (a), short (b), and load (c), with data-based calibration method applied at cryogenic temperatures and compared with room temperature (RT) measurement using ECal. Blue and green lines are the response of data-based calibration and the orange and pink lines are the response of ECal. The phase response of the short for both calibrations are almost identical so they are not distinguishable in plots (a) and (b). Assuming $2/3~c$ phase velocity, the short has a physical length of $2~$mm and the open is about $1.72~$mm long. For the load measurement, the signals are at very low levels due to the low-reflective nature of the standard; thus, the phase responses contain no useful information and are not shown here.
   }
    \label{fig:CalibrationStandards}
\end{figure}

In this section, a calibration method using data-based description evaluated at ambient and cryogenic temperatures was applied to the measurement of the calibration standards, and the results are used as first-level verification. Preliminary ambient temperature calibrations using electronic calibration (ECal) were conducted as a baseline and gave expected results (see Fig.~\ref{fig:CalibrationStandards}). 

At cryogenic temperatures, the short- and open-calibration samples are expected to demonstrate similar microwave responses as  ambient, while the load is expected to deviate slightly. 

Results of these tests are shown in Fig.~\ref{fig:CalibrationStandards}. With the data-based calibration we find that the return loss is within $0\pm0.01~$dB for the open and $-0.02\pm0.01~$dB for the short. Both short and open have linear phase delay indicating uniform offsets. Compared to ambient Ecal results, we can see that for open and short, the magnitudes are both very close to $0~$dB and the cryogenic one has lower loss, which is expected. The phase responses are also almost the same, indicating very small length change: the open becomes 4\% shorter whereas the short has no obvious change at cryogenic temperature. For the load, at $15~$mK, it has a flatter response than the room temperature one while on average they both have more than $30~$dB isolation over the calibration bandwidth. Overall, the data-based calibration at cryogenic temperature gives results that agree with expectations.

\begin{figure}
   \centering
   \includegraphics[width=\linewidth]{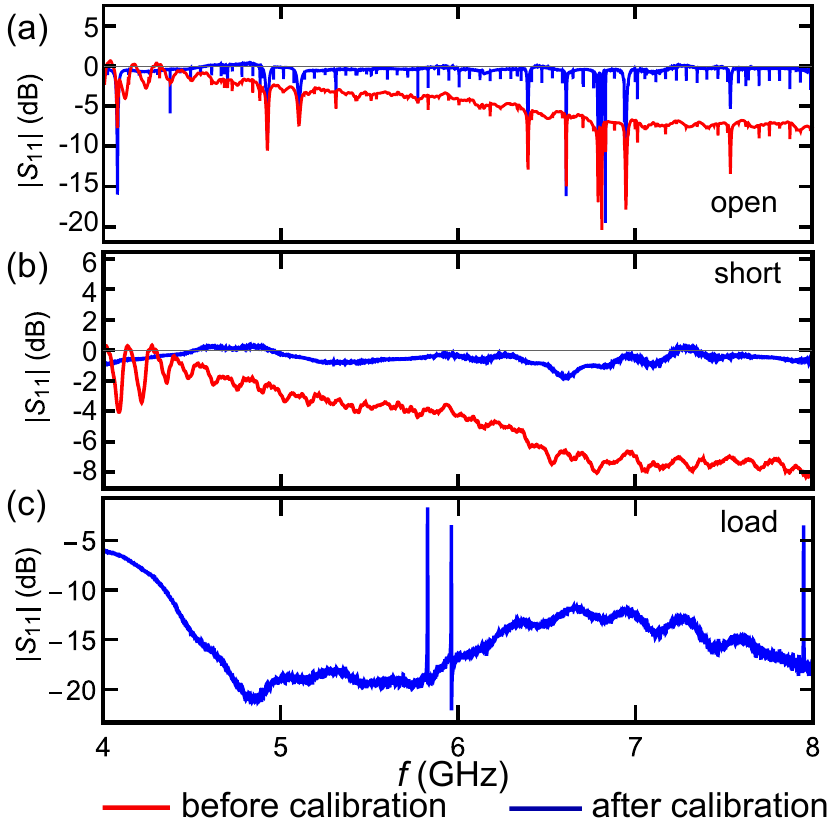}
   \caption{(Colored Online) Measurement results of three 2D superconducting resonators at $T_{base}$ with data-based calibration applied. Samples are installed in similar sample boxes with unused ports terminated by standards. Panel (a) shows a half-wave resonator sample with fundamental resonance frequency of $59.5~$MHz terminated by an open. The resonances are over-coupled with $Q\sim10^4$, which appear as deep and broad dips. Panel (b) shows a sample with eight resonators coupled to an on-chip coplanar waveguide (CPW) feedline in hanger-mode terminated by a short. The resonances are critically-coupled with $Q\sim10^6$, which are too narrow to be seen here. Panel (c) shows a sample with three resonators coupled to an on-chip feedline in hanger-mode terminated by a load. The resonances are over-coupled with $Q\sim10^5$, which appear as peaks. 
}
\label{fig:VerificationLines}
\end{figure}

The next level of verification is to measure the test standards through planar superconducting resonators mounted in sample boxes. Because the measurement chain goes through wirebonds, sample boxes, resonators, and SMA connectors, we expect to observe ripples in the frequency response due to impedance mismatches.
In Fig.~\ref{fig:VerificationLines}, we show calibrations applied on: a half-wave resonator terminated by a test open (a), an on-chip transmission feedline with eight high $Q$ resonators coupled to it in hanger-mode terminated by a test short (b),  and an on-chip transmission feedline coupled with three low-$Q$ resonators in hanger-mode terminated by a load (c). For the test open and short, we also compare the results to the un-calibrated measurement. For the samples terminated by short and open, the data-based calibration results have a mostly flat background with small ripples. For the load, the much higher return loss compared to a bare load measurement [Fig.~\ref{fig:CalibrationStandards}(c)] reveals the mismatch from wirebonds, SMA connectors, and sample boxes. We have noticed that the after-calibration baseline can slightly pass $0~$dB at certain frequencies. This indicates the calibration is not perfect.

\begin{figure}
   \centering
   \includegraphics[width=\linewidth]{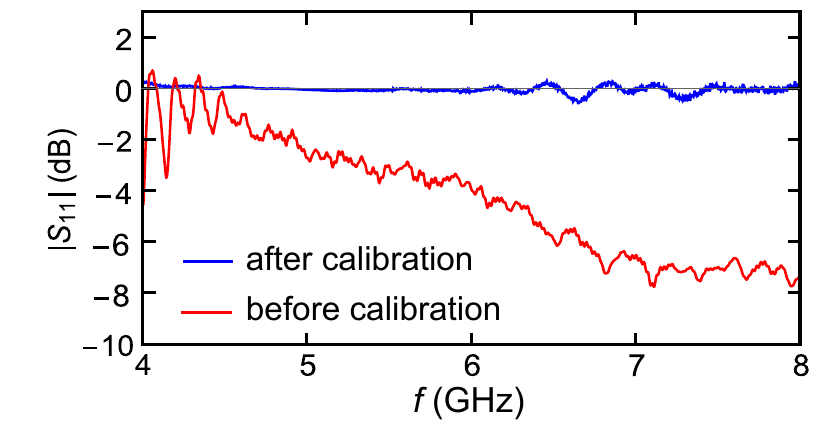}
   \caption{(Colored Online) Comparison of the magnitude of reflection measurement of a 3D cavity input port with and without data-based calibration applied.
   }
    \label{fig:LLNL3DCalibrated}
\end{figure}

Since 3D superconducting cavities have high reflection off-resonance and ultra low loss on-resonance, they are especially useful to verify the calibration. Fig.~\ref{fig:LLNL3DCalibrated} illustrates the effect of calibration on the background of a 3D cavity with four quarter wave stub resonators inside (see Fig.~\ref{fig:3D_cavity}). $Q_i$ extraction is discussed in Sec.~\ref{sec:Cal_Applied_to_Reflection}. The data-based calibration successfully removes the large ripples at low frequencies, the frequency dependent loss which is associated with cables and filters, and small ripples across the entire experimental frequency range. Some residual deviations of $<1~$dB remain between $6.5~\mathrm{GHz}$ and $7.5~\mathrm{GHz}$ and can be explained by the mismatch from the launching SMA connector of the cavity.   

Based on the above three levels of verification, we can conclude that, although the calibration standards are expected to experience small changes in microwave behavior between ambient and cryogenic temperatures, data-based calibration results agree with measurement expectations. The three terms of the error adapter can be found in Appendix~\ref{sec:error_box}.

\section{Resonator Measurement with Calibration}\label{sec:Cal_Applied_to_Reflection}

We measure one 3D superconducting cavity and one 2D planar resonator with (reference plane on the sample input) and without (reference plane outside of DR) data-based calibration to demonstrate the effect on performance metric $Q_i$. Since the $Q_i$ of overcoupled resonators is more strongly affected by errors due to impedance mismatch,~\citep{Khalil2012} overcoupled resonators are measured in this experiment. The data processing codebase and fitting routines are available online.~\footnote{https://github.com/Boulder-Cryogenic-Quantum-Testbed/measurement/}

\subsection{3D Cavity Measurements}\label{sec:3D cavity measurement}

The objective of cryogenic calibration is to measure cryogenic devices with higher accuracy than the measurement with reference plane outside of DR. To demonstrate this we measured a 3D cavity with four stub resonators (see Appendix~\ref{sec:3d_cavity} for details), referred to as resonators 1-4, and designed to have fundamental resonance frequencies $3.80~$GHz, $4.04~$GHz, $4.09~$GHz and $4.27~$GHz. It is the same device used in Sec~\ref{sec:verfication} with background measured (Fig.~\ref{fig:LLNL3DCalibrated}) for calibration verification .
We measure this device in reflection through one of the two ports, strengthening the coupling to the two resonators nearest to the measured port (1 and 3) and reducing the coupling to the further two resonators (2 and 4).
The measured resonance frequencies for the four resonators are $3.85~$GHz, $4.02~$GHz, $4.11~$GHz and $4.25~$GHz, close to the designed values. $Q_c$ values are extracted to be $5.8\times10^5$, $1.8\times10^6$, $4.3\times10^5$, and $3.7\times10^6$ respectively. 

\begin{figure}
\includegraphics[width=\linewidth]{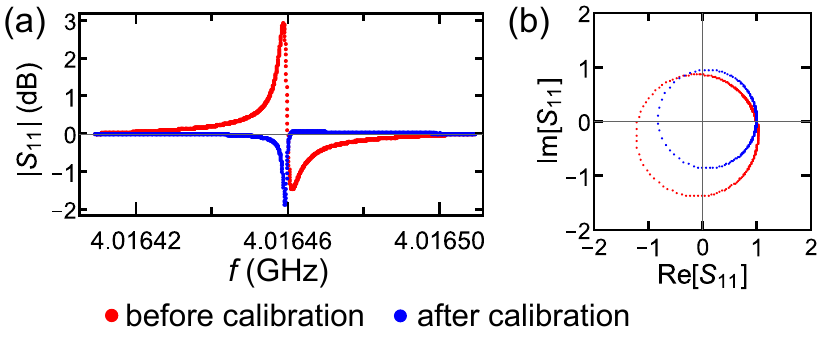}
\caption{\label{fig:Res2Res3} (Colored online) Measurements of resonator 2 of the 3D cavity ($4.02~\mathrm{GHz}$) before (in red) and after (in blue) calibration at $-85~$dBm applied power. (a) is the magnitude and (b) is the resonance circle on complex plane after normalization.}
\end{figure}

A detailed comparison of calibrated and uncalibrated $S_{21}$ response for 3D stub resonator 2 is illustrated in Fig.~\ref{fig:Res2Res3}. Before correction, the magnitude line-shape is asymmetric and the resonance circle has a rotation of $\theta=0.23$. The corrected line-shape is closer to a symmetric Lorentzian, with $\theta=-0.05$.

\begin{figure}[t]
\includegraphics[width=\linewidth]{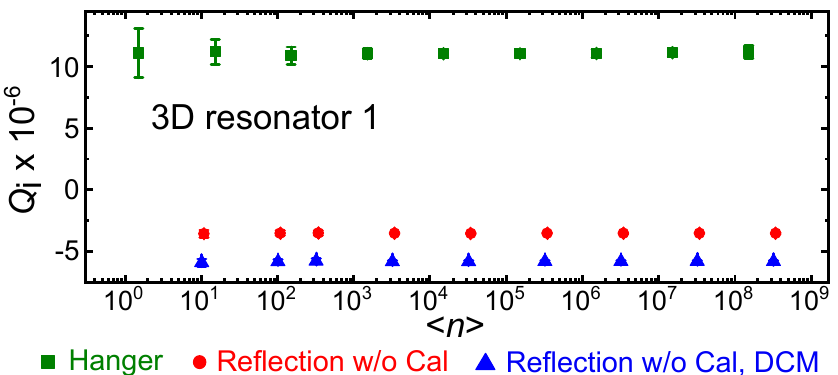}
\caption{\label{fig:Res1_3D} (Colored online) Linear-log scale plot of the $Q_i$ measurements of resonator 1 of the 3D cavity in reflection and in hanger-mode using an SMA tee (see appendix.~\ref{sec:tee} for the background measurement), with hanger-mode measurements serving as benchmarks. Green squares is hanger measurement; Red circles is reflection measurement without calibration applied; Blue triangles is the reflection measurement using DCM.
}
\end{figure}

\begin{figure}
\includegraphics[width=\linewidth]{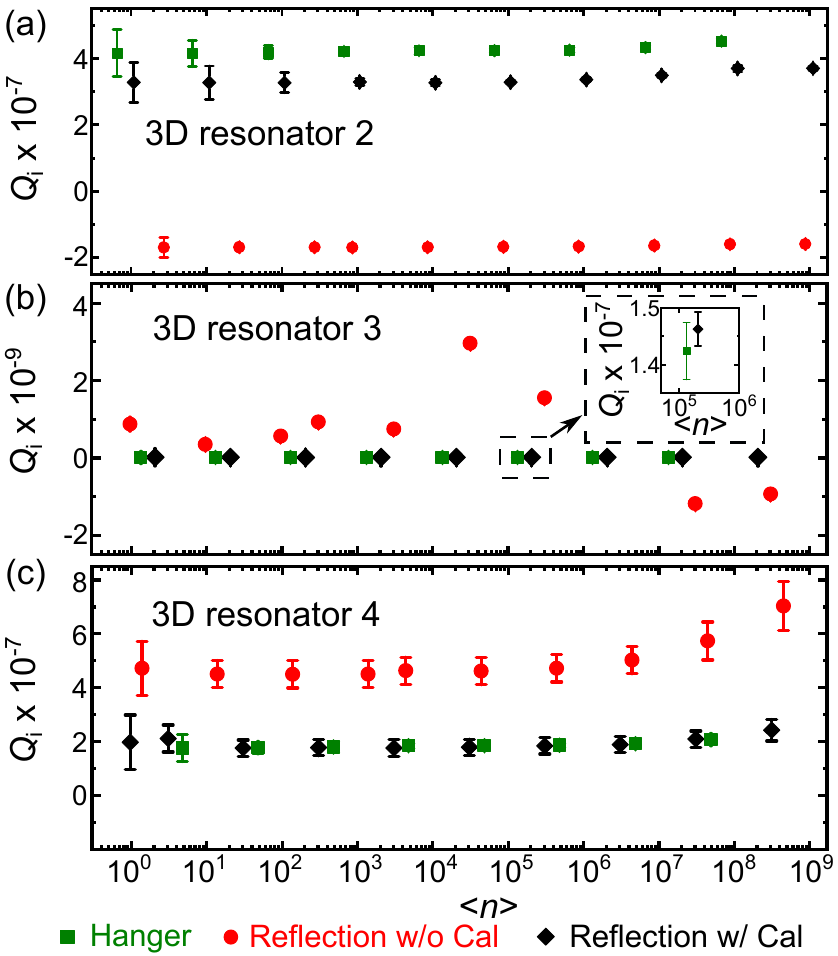}
\caption{\label{fig:3D_Qi} (Colored online) Linear-log scale plot of the internal quality factor $Q_i$ as a function of average photon number in the cavity $\langle n \rangle$ for three of the four quarter-wave resonators inside a 3D cavity. Plot (a), (b) and (c) are results of the 3D stub resonators 2, 3 and 4 respectively. Inset in (b) shows the hanger and calibrated reflection results zoomed in near $\langle n \rangle= 10^5 - 10^6$. Green squares is hanger measurement; Red circles is reflection measurement without calibration applied; Black diamonds is reflection measurement with calibration applied.
} 
\end{figure}
Power sweep results are shown in Figs.~\ref{fig:Res1_3D} and~\ref{fig:3D_Qi}. The power is converted to average photon number in the cavity $\langle n \rangle$, determined by 
\citep{GaoThesis,Bruno2015}
\begin{equation}
\langle n \rangle = \frac{2}{\hbar \omega_0^2} \frac{Z_0}{Z_r} \frac{Q^2}{Q_c} P_{\mathrm{app}},
\end{equation} with the applied power ($P_{app}$) estimated by the VNA output power subtracting $70~$dB cable loss and attenuation, and $Z_0, Z_r$ are the characteristic impedance of the environment and the resonator impedance respectively. Resonator 1 cannot be calibrated since the frequency $f_0$ = $3.85~$GHz is outside the calibration bandwidth, but it can be used as an example to demonstrate that DCM does not apply to the reflection measurement (shown in the Fig.~\ref{fig:Res1_3D}). The $Q_i$ values reported in Fig.~\ref{fig:3D_Qi} are extracted using Eqs.~\ref{eq:Qi original} and~\ref{eq:S21 reflection_rotation}. Resonators 2 and 3 [Fig.~\ref{fig:3D_Qi}(a)-(b)] have unphysical negative $Q_i$ before calibration, indicating a strong impedance mismatch due to wiring and microwave components. After calibration, $Q_i$ is within the expected range of $1.5\times 10^{7} - 4\times10^7$. 
Fig.~\ref{fig:3D_Qi} demonstrates a comparison of calibrated and uncalibrated reflection measurements to benchmark hanger-mode measurements, which are unaffected by choice of reference plane. For 3D stub resonators 2-4 the calibration moves reflection $Q_i$ closer to the values when the resonators are measured in the self-referencing hanger-mode.

\subsection{2D Planar Resonator Measurements}
\label{sec:2D measurement}

\begin{figure}
\includegraphics[width=\linewidth]{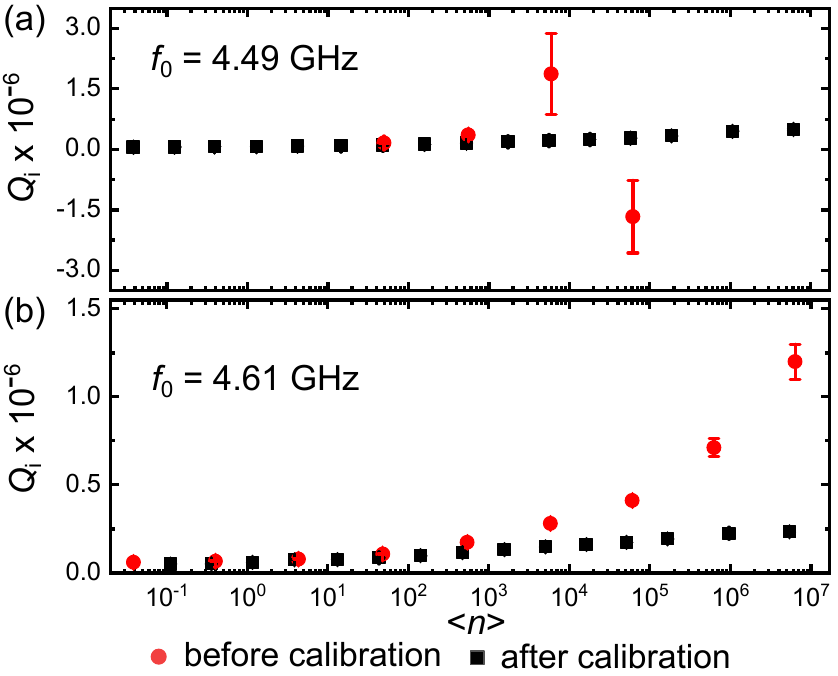}
\caption{\label{fig:2DQi} (Colored online) Linear-log scale plot of the internal quality factor $Q_i$ as a function of average photon number in the cavity $\langle n \rangle$ for two high-order, overcoupled harmonic modes of a 2D half-wave resonator at $4.49~\mathrm{GHz}$ (a) and $4.61~\mathrm{GHz}$ (b) in reflection. 
}
\end{figure}

The 2D resonator sample measured is a $40~$cm long half-wave coplanar waveguide (CPW) resonator made of NbTiN, with a design as in Ref.~\citenum{erickson2014frequency} and the far end port terminated by a short. It is the same sample mentioned in Sec~\ref{sec:verfication} with the far end terminated by an open and then measured for calibration verification. We measure resonances at $4.49~$GHz and $4.61~$GHz. The resonator mode spacing is around 59.5 MHz, close to the designed value of $60~\mathrm{MHz}$. The resonator is designed to be near critically coupled at its fundamental mode, but it is strongly overcoupled at high-order harmonics in the $4-8~\mathrm{GHz}$ measuring frequency bandwidth. The $Q_c$ is around $1.7\times10^4$ for both modes.

Fig.~\ref{fig:2DQi} shows the $Q_i$ of this device before and after calibration as a function of power. We observe both ultra-high and negative $Q_i$ values before calibration. After calibration, $Q_i$ is positive at all powers and an S-shaped curve is seen (see Appendix~\ref{sec:TLS loss}), characteristic of TLS power dependence.~\citep{Richardson2016,kalacheva2020improving} We extract an intrinsic TLS loss of $F \tan \delta_0= (1.8\pm 0.11)\times 10^{-5}$ for the $4.49~$GHz resonance and $F \tan \delta_0= (1.74\pm0.09)\times 10^{-5}$ for the $4.61~$GHz resonance. As the power decreases and $Q_i$ approaches the critically coupled regime for the $4.61~$GHz resonance, the difference between before and after calibration becomes smaller. This agrees with our expectation that the error in $Q_i$ extraction drops with decreasing $Q_i/Q_c$ ratio.

\section{Conclusions and Future Work}\label{sec:conclusion}
To conclude, we have demonstrated a  single-port, data-based SOL calibration at cryogenic temperature. This was achieved by accessing VNA's front panel configuration so the calibration can be performed using the built-in calibration functions.
This data-based calibration has high accuracy and can be easily adapted for 2D and 3D superconducting microwave resonator measurements in reflection. We demonstrate that DCM does not correct rotations in reflection measurements, and therefore calibration is essential for those experiments. 

The data-based SOL calibration with VNA reconfiguration described in this paper can be easily extended to a two-port setup by implementing a second set of standards and a cable that serves as an unknown through. Future work includes implementing this two-port data-based short-open-load-through (SOLT) calibration following the same procedure, as well as other two-port methods.

\begin{acknowledgments}
We wish to acknowledge David Fork for his tremendous help setting up the lab and equipment; Dylan Williams for useful discussions; Evan Jeffrey for discussions about measuring resonators in reflection; Kevin Chaves and Jonathan DuBois for designing and fabricating the 3D cavity; Michael Vissers for fabricating the 2D resonator; Xioranny Linares and Sonali Sarpotdar for administrative support; the partial support of the Army Research Office, Fermi National Laboratory, Google, the NIST Quantum Initiative, the U.S. Department of Energy and Lawrence Livermore National Laboratory under Contract DE-AC52-07NA27344 and the Laboratory Directed Research and Development grant 19-ERD-013, and the National Science Foundation Grant Number 1839136 and the National Science Foundation (Award No.  1839136). 
\end{acknowledgments}

\appendix


\section{The resonator model and diameter correction}
\label{sec:resonator diameter correction}

\begin{figure}[htb!]
\includegraphics[width=\linewidth]{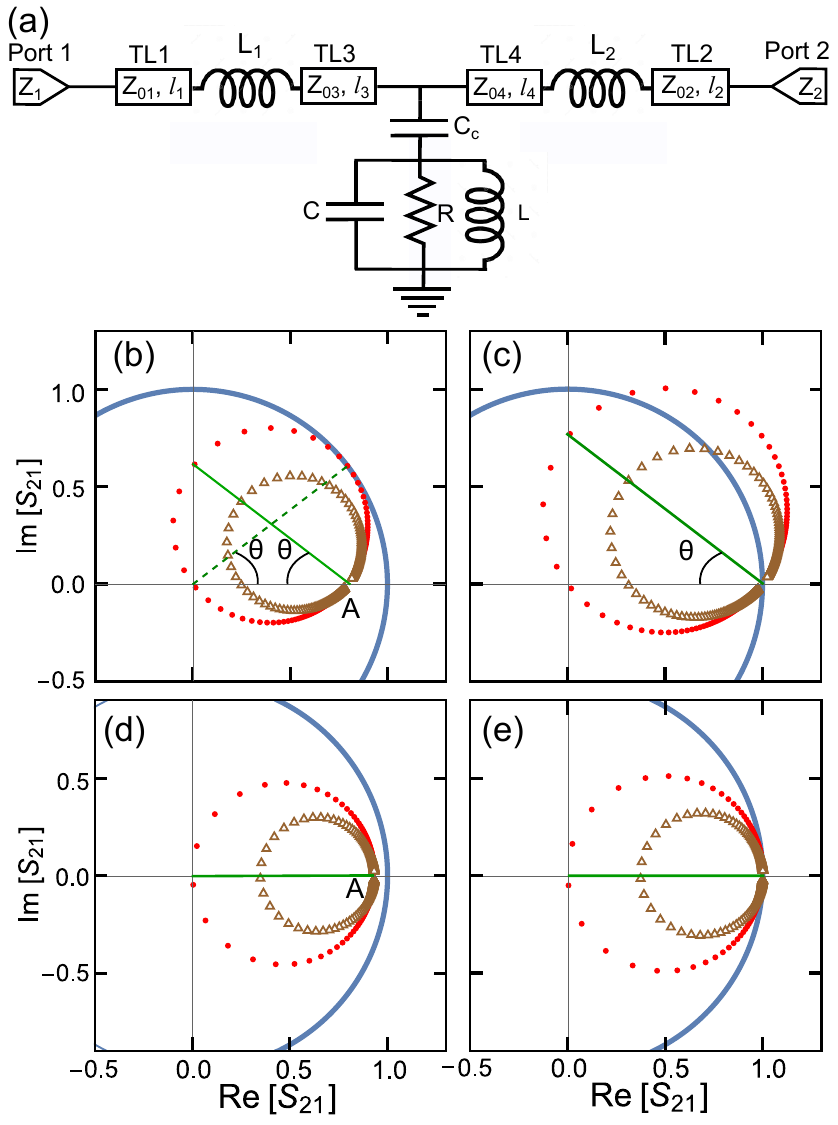}
\caption{\label{fig:HangerDCM} (Colored online) Simulated $S_{21}$ on the complex plane of a parallel LCR resonator connected in hanger-mode with wirebonds represented by inductors L$_1$ and L$_2$ on either side. (a) The circuit diagram of simulation setup, similar to Fig.~\ref{fig:circuit_diagram}(a). (b)[(d)] is the resonance circles before normalization while (c)[(e)] is the corresponding circles after normalization. The simulation parameters are: $Z_{1,2}=Z_{01,02,03,04}=50~\Omega$, $L_1=0.5~$nH, $L_2=1.5~$nH, $C=580~$fF, $L=1.2~$nH, $R=10^{12}~\Omega$ for red solid dot and $10^8~\Omega$ for brown open triangle,
$C_c=1~$fF sets $Q_c$ to be approximately $6\times10^5$. $l_3=l_4=90^\circ$ for (b) and $l_3=107^\circ$ and $l_4=17^\circ$ for (d). The off-resonance point is marked as A and $\theta$ is the rotation angel. The blue circle is the unity circle.
}
\end{figure}

Here we briefly discuss the mismatch in hanger-mode and how DCM works by reviewing the resonance circle. Consider a parallel LCR resonator capacitively coupled to a transmission feedline with wirebonds connecting to the environment as shown in Fig.~\ref{fig:HangerDCM}(a). This is similar to the actual measurement setup if we neglect the attenuation, amplification and cable loss. The wirebonds can be treated as inductors. To simplify the discussion, here we assume the mismatch only comes from the wirebonds, meaning all the port impedance and transmission line characteristic impedance are the same ($50~\Omega$), and only the wirebonds have non-zero inductance.

The mismatch will suppress the background level to be lower than 1. For a high $Q$ resonance, the background is nearly at a constant level for several bandwidths and can be noted as $A$. If $Q$ is too low, $A$ is frequency-dependent. When this dependency is not linear, the resonance circle will be distorted and cannot be used to fit for $Q_i$. Therefore, avoiding mismatch is still desirable. The suppression of background from mismatch is indistinguishable from having loss or attenuation in the setup. As a result, the normalization process will magnify $A$ to be 1, thus magnify the size of the resonance circle. Shown in Fig.~\ref{fig:HangerDCM}(c), some area passes unity after normalization, which appears to get gain.

For a strongly overcoupled resonator (near infinite $Q_i$) in hanger-mode, near resonance, the impedance is approaching zero, thus acts as a short to ground which reflects all the incoming signal back. Therefore, the measured the transmission $S_{21}$ is 0. This means the resonance circle has to pass through the origin, which is reflected by the red circle in Fig.~\ref{fig:HangerDCM}(b) and (d). When the rotation angle $\theta$ is not zero, this circle also intersects the imaginary axis at another point. It's not hard to find that this intersection is the on-resonance point. The diameter of the circle is $A/\cos \theta$. The circle cannot pass beyond the unity circle due to energy conservation, thus, as illustrated in Fig.~\ref{fig:HangerDCM}(b), the maximum allowed rotation $\theta_{max}$ happens when the resonance circle and the unity circle are tangent: 
\begin{equation}
\label{eq:theta_max}
    \theta_{max}=\cos^{-1}A.
\end{equation}
For a given off-resonance transmission level $A$, $\theta$ is limited to the range of $[-\theta_{max},\theta_{max}]$. When $\theta$ is equal to 0, it does not imply no impedance mismatch. As shown in Fig.~\ref{fig:HangerDCM}(d), $A$ is about 0.9, which is smaller than 1, indicating imperfect background transmission, while the rotation is very small ($\theta=0.02$).

From Fig.~\ref{fig:HangerDCM}(c), it can be seen that the diameter of the red circle after normalization is $1/\cos \theta$. However, for a resonance with infinite $Q_i$, $Q/Q_c$ should be equal to 1. This implies that the diameter is magnified by $1/\cos \theta$. $Q_c$ extracted from Eq.~\ref{eq:Qi original} will be lowered the actual value and $Q_i$ will be raised. To correct this error, $Q_c$ needs to be corrected as $Q_c/\cos\theta$ and then $Q_i$ can be found through Eq.~\ref{eq:Qi correct}. In other words, normalizing the off-resonance to 1 can cause error in $Q_i$ extraction because the diameter gets magnified. However, the amount of magnification is a single value function of $\theta$. This is the reason for hanger-mode, the impedance induced error can be corrected through the amount of rotation of the resonance circle. The circle for finite $Q_i$ (brown triangle) has the same amount of magnification in the diameter, thus can also be corrected using DCM. Following the same process of reviewing the resonance circle, it can be seen that the diameter magnification and rotation angle $\theta$ in both reflection- and transmission-mode are not correlated as a single value function. As a result, both modes require a calibration to accurately measure $Q_i$.~\citep{Probst2015,wang2019mode}

\section{3D cavity}
\label{sec:3d_cavity}

\begin{figure}
\includegraphics[width=\linewidth]{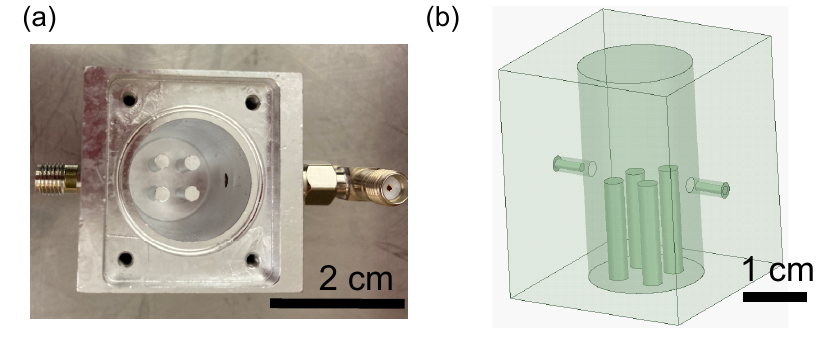}
\caption{\label{fig:3D_cavity} (Colored online) Top-down view of the cavity before assembly (a) and the cavity design (b). The device is a 3D RF cylindrical cavity encompassing four quarter-wave stub resonators of varying heights. 
}
\end{figure}

The 3D cavity (shown in Fig.~\ref{fig:3D_cavity}) was conventionally machined using a computer numerical control (CNC) machine and is comprised of a single piece of 5N5 high purity aluminum. Surfaces were treated in a bath of commercially available phosphoric-nitric acid mix at $50^{\circ}$C for four hours while refreshing at the two-hour point to prevent etchant saturation. This process removed 100 $\mu$m of material from the inner surface of the cavity in order to improve surface quality thereby enhancing the internal quality factor of the resonator modes. There are two coupling ports located on opposite sides of the outer rectangular walls which allow us to evanescently couple to the resonator modes via SMA pin couplers.

\section{TLS fitting of 3D and 2D resonator}\label{sec:TLS loss}

\begin{figure}[b!]
\includegraphics[width=\linewidth]{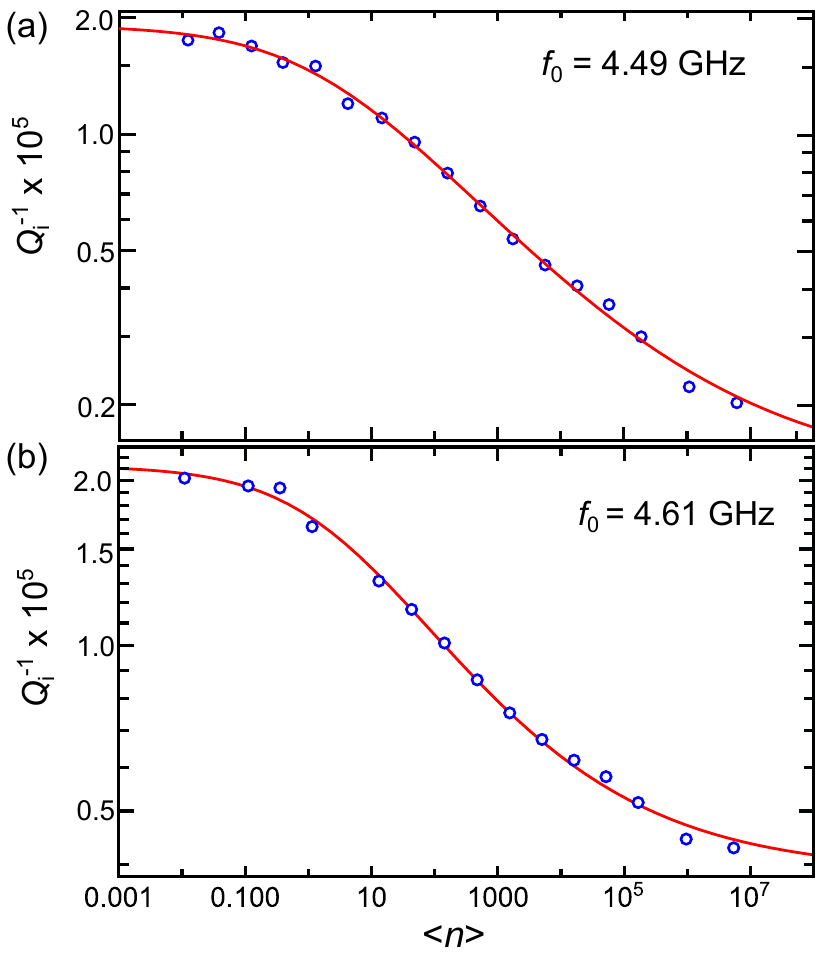}
\caption{\label{fig:TLS} (Colored online) Logarithmic plot of the internal loss 1/$Q_i$ as a function of photon number $\langle n \rangle$ of $4.49~$GHz resonance (a) and $4.61~$GHz resonance (b) of the 2D resonator measured in Sec.~\ref{sec:2D measurement} in blue open circles. The red lines are the fitted two-level system (TLS) loss extraction function.
}
\end{figure}

The power-dependent internal loss $1/Q_i$ of a resonator can be represented by~\citep{Pappas2011,Burnett2018}
\begin{equation}
\label{eq:TLS_Loss}
    \frac{1}{Q_i}=F\frac{1}{Q_{i,0}}\frac{ \tanh \left(\frac{h f_0}{2 k_\mathrm{B} T}\right)}{\sqrt{1+\left(\frac{\langle n \rangle}{n_c}\right)^{\beta}}}+\frac{1}{Q_{i,\mathrm{other}}},
\end{equation}
where $F$ is the filling factor, $1/Q_{i,0}$ is the intrinsic TLS loss, $1/Q_{i,\mathrm{other}}$ is the power-independent loss, $\langle n \rangle$ is the average photon number inside the resonator, and $n_c$ is the critical photon number where the TLS loss starts to saturate. The fitting results of the $4.49~$GHz and $4.61~$GHz resonances can be found in Tab.~\ref{tab:table2} and Fig.~\ref{fig:TLS} shows the loss as a function of photon number for the measured data and fitted function. The intrinsic TLS loss for these two resonances are similar, as expected.

\begin{table}
\caption{\label{tab:table2}TLS fitting results of the 2D resonator}
\begin{ruledtabular}
\begin{tabular}{cccc}
$f_0~$(GHz)  & F/$Q_{i,0} \times 10^{5}$  & $n_c$ &   $Q_{i,\mathrm{other}} \times 10^{-5}$  \\
 \hline

4.49   & $1.80 \pm 0.11$  & $1.74 \pm 0.74$  &   $7.7 \pm 0.29$  \\
4.61    & $1.74 \pm 0.09$  & $2.17 \pm 0.81$  &   $2.57 \pm 0.26$ 
\end{tabular}
\end{ruledtabular}
\end{table}

\section{Switch repeatability and stability }\label{sec:switch}

\begin{figure}
\includegraphics[width=\linewidth]{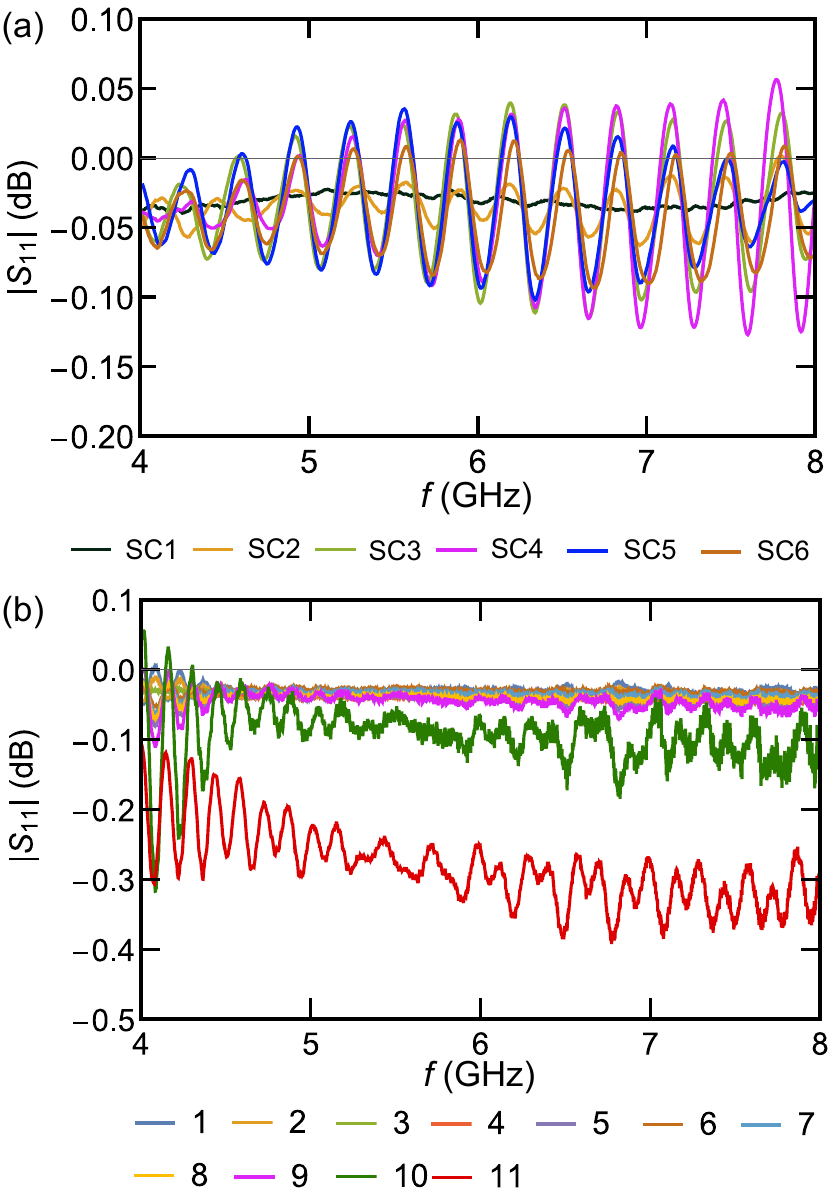}
\caption{\label{fig:switch_test} (Colored online) (a) Measurement of the standard short at different switch channels at ambient temperature with ECal performed at switch channel 1 and applied to each measurement. ``SC" in the legend standards for ``switch channel". (b) Repeated measurement of the standard short using the data-based calibration 11 times with the switch randomly opened and closed or switched to different position and back over a $24~$hour period at $15~$mK. The number in the legend is the index of the corresponding measurement. 
}
\end{figure}

In Sec.~\ref{sec:experiment_Setup}, we assume the switch channels are identical. To test this assumption, we make series of measurements at ambient temperature and assume that differences here will also be present at cryogenic temperatures and vice versa. We connect the same cables used in the cryogenic temperature calibration to all six switch ports, perform electronic calibration (ECal) using the calibration set from the manufacturer, and then measure the standard short. For example, as shown in Fig.~\ref{fig:switch_test}(a), an ECal is performed on channel 1. We then measure the standard short using switch channel 1 in reflection with ECal applied and use the result as benchmark. We then manually connect and disconnect the standard short to each one of the remaining channels, measure it in reflection and compare the results to channel 1 result. We then perform ECal on channel 2 and repeat the same procedure. We find that the differences between channels are within $\pm 0.1~$dB. This is similar to what other works have found,~\citep{ranzani2013two,yeh2013situ} and is small enough to safely assume all channels are identical. 

We also test the switch stability at cryogenic temperature. We use the data-based calibration to measure the standard short on channel 5 (see Fig.~\ref{fig:fridge_wiring}). We then randomly connect and disconnect or switch back and forth to other channels. After each switch operation, we wait for about $30~$minutes for mixing chamber temperature stabilization. 11 measurements were performed within a $24~$hour period. As shown in Fig.~\ref{fig:switch_test}(b), drift continues to increase over this period, with a total drift of around $0.25~$dB from the first measurement. For the first 9 measurements, which spans about 5 hours, the drift is below $0.05~$dB. Based on the test results, we recommend performing calibration daily.

\section{Background measurement of SMA Tee at ambient temperature}\label{sec:tee}
\begin{figure}
\includegraphics[width=\linewidth]{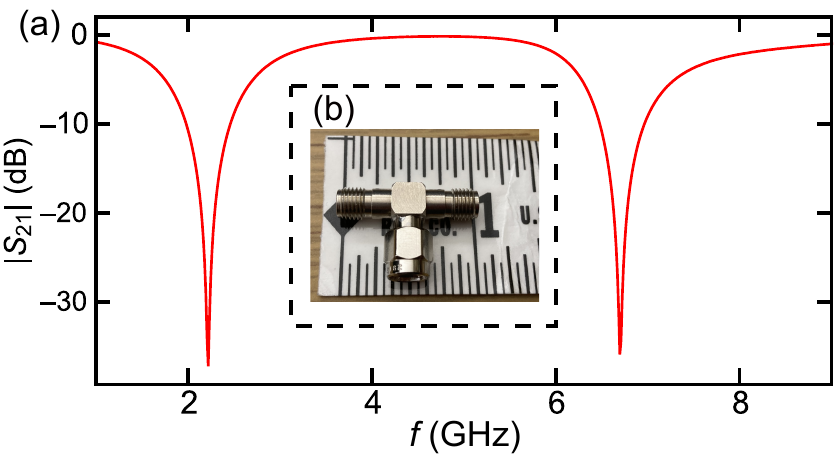}
\caption{\label{fig:tee} (Colored online) (a) The background transmission of the SMA Tee used for hanger-mode measurement of the 3D cavity described in Sec.~\ref{sec:3D cavity measurement} at ambient temperature from $1~$GHz to $9~$GHz. The SMA tee introduces a mismatch that causes large dips in transmission near $2.1~$GHz and $6.8~$GHz. However, for the frequency range of interest in this work, near $4~$GHz, transmission is nearly flat with minimal suppression from $0~$dB. (b) The actual SMA Tee used in this work.
}
\end{figure}

Using a 3-port device like the SMA tee in hanger mode inevitably introduces mismatches. We measured the transmission of the Tee used for hanger-mode at ambient temperature to make sure the mismatch does not have large effect at the frequency range of the resonators. In Fig.~\ref{fig:tee}, we can see the near $4~$GHz, the background is almost flat while near $6.8~$GHz there is a large dip. Therefore, using this tee will not distort the line-shapes nor lower the signal for the 3D cavity resonances. But if the designed frequencies are in the range of $6.5~$GHz to $7.5~$GHz, the tee can increase the difficulty of hanger-mode measurements. 

\section{Error adapter}\label{sec:error_box}

\begin{figure}
\includegraphics[width=\linewidth]{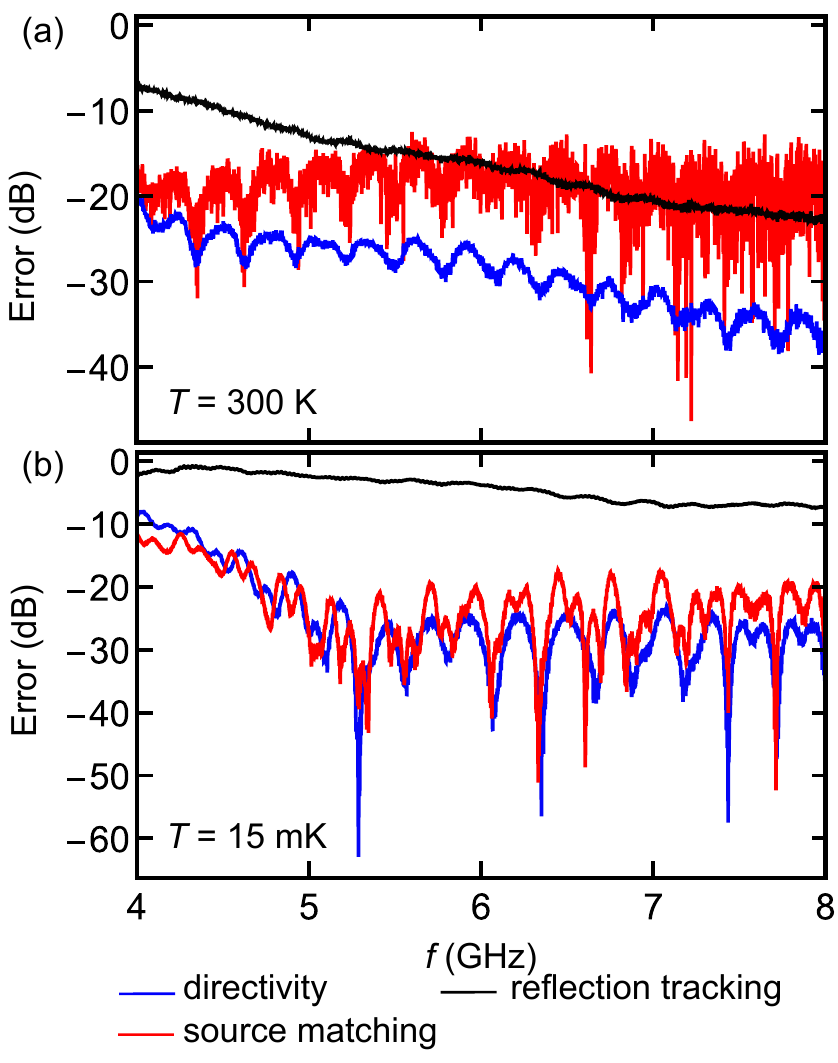}
\caption{\label{fig:Error_box} (Colored online) The error adapter terms of the data-based cryogenic calibration. (a) is the RT ECal results and (b) is the 15 mK data-based results.
}
\end{figure}

The three terms of the error adapter for the data-based cryogenic calibration are shown in Fig.~\ref{fig:Error_box}. The biggest change we expect when the fridge is cold versus warm would be the superconducting cables connecting $3~$K and $15~$mK plates having lower loss and better match. This is shown by the overall lower error extraction [Fig~\ref{fig:Error_box}(b)]. 
The overall source match error extracted by the cryogenic calibration is nearly is about -30 dB, which is less than 0.1$\%$, in the region from 5 to $8~$GHz. The directivity, which drops from -20 to about $-30~$dB, is improved due to the improvement of isolation of the reference and test signals. For the frequency range below $5~$GHz, one possible explanation is that the change is due to the attenuators' response and requires further investigation. The reflection tracking, which is a frequency response performance, is improved from $-10~$dB to $-20~$dB at ambient temperature to less than $-10~$dB at cryogenic temperature as well. In conclusion, we believe moving the reference plane to the operating temperature environment is necessary to get better error extraction.

\bibliography{OnePortCal_Physics}

\providecommand{\noopsort}[1]{}\providecommand{\singleletter}[1]{#1}
\begin{thebibliography}{46}%
\makeatletter
\providecommand \@ifxundefined [1]{%
 \@ifx{#1\undefined}
}%
\providecommand \@ifnum [1]{%
 \ifnum #1\expandafter \@firstoftwo
 \else \expandafter \@secondoftwo
 \fi
}%
\providecommand \@ifx [1]{%
 \ifx #1\expandafter \@firstoftwo
 \else \expandafter \@secondoftwo
 \fi
}%
\providecommand \natexlab [1]{#1}%
\providecommand \enquote  [1]{``#1''}%
\providecommand \bibnamefont  [1]{#1}%
\providecommand \bibfnamefont [1]{#1}%
\providecommand \citenamefont [1]{#1}%
\providecommand \href@noop [0]{\@secondoftwo}%
\providecommand \href [0]{\begingroup \@sanitize@url \@href}%
\providecommand \@href[1]{\@@startlink{#1}\@@href}%
\providecommand \@@href[1]{\endgroup#1\@@endlink}%
\providecommand \@sanitize@url [0]{\catcode `\\12\catcode `\$12\catcode
  `\&12\catcode `\#12\catcode `\^12\catcode `\_12\catcode `\%12\relax}%
\providecommand \@@startlink[1]{}%
\providecommand \@@endlink[0]{}%
\providecommand \url  [0]{\begingroup\@sanitize@url \@url }%
\providecommand \@url [1]{\endgroup\@href {#1}{\urlprefix }}%
\providecommand \urlprefix  [0]{URL }%
\providecommand \Eprint [0]{\href }%
\providecommand \doibase [0]{http://dx.doi.org/}%
\providecommand \selectlanguage [0]{\@gobble}%
\providecommand \bibinfo  [0]{\@secondoftwo}%
\providecommand \bibfield  [0]{\@secondoftwo}%
\providecommand \translation [1]{[#1]}%
\providecommand \BibitemOpen [0]{}%
\providecommand \bibitemStop [0]{}%
\providecommand \bibitemNoStop [0]{.\EOS\space}%
\providecommand \EOS [0]{\spacefactor3000\relax}%
\providecommand \BibitemShut  [1]{\csname bibitem#1\endcsname}%
\let\auto@bib@innerbib\@empty
\bibitem [{\citenamefont {Sete}, \citenamefont {Martinis},\ and\ \citenamefont
  {Korotkov}(2015)}]{sete2015quantum}%
  \BibitemOpen
  \bibfield  {author} {\bibinfo {author} {\bibfnamefont {E.~A.}\ \bibnamefont
  {Sete}}, \bibinfo {author} {\bibfnamefont {J.~M.}\ \bibnamefont {Martinis}},
  \ and\ \bibinfo {author} {\bibfnamefont {A.~N.}\ \bibnamefont {Korotkov}},\
  }\href@noop {} {\bibfield  {journal} {\bibinfo  {journal} {Phys. Rev. A}\
  }\textbf {\bibinfo {volume} {92}},\ \bibinfo {pages} {012325} (\bibinfo
  {year} {2015})}\BibitemShut {NoStop}%
\bibitem [{\citenamefont {Reed}\ \emph {et~al.}(2010)\citenamefont {Reed},
  \citenamefont {Johnson}, \citenamefont {Houck}, \citenamefont {DiCarlo},
  \citenamefont {Chow}, \citenamefont {Schuster}, \citenamefont {Frunzio},\
  and\ \citenamefont {Schoelkopf}}]{reed2010fast}%
  \BibitemOpen
  \bibfield  {author} {\bibinfo {author} {\bibfnamefont {M.~D.}\ \bibnamefont
  {Reed}}, \bibinfo {author} {\bibfnamefont {B.~R.}\ \bibnamefont {Johnson}},
  \bibinfo {author} {\bibfnamefont {A.~A.}\ \bibnamefont {Houck}}, \bibinfo
  {author} {\bibfnamefont {L.}~\bibnamefont {DiCarlo}}, \bibinfo {author}
  {\bibfnamefont {J.~M.}\ \bibnamefont {Chow}}, \bibinfo {author}
  {\bibfnamefont {D.~I.}\ \bibnamefont {Schuster}}, \bibinfo {author}
  {\bibfnamefont {L.}~\bibnamefont {Frunzio}}, \ and\ \bibinfo {author}
  {\bibfnamefont {R.~J.}\ \bibnamefont {Schoelkopf}},\ }\href@noop {}
  {\bibfield  {journal} {\bibinfo  {journal} {Appl. Phys. Lett.}\ }\textbf
  {\bibinfo {volume} {96}},\ \bibinfo {pages} {203110} (\bibinfo {year}
  {2010})}\BibitemShut {NoStop}%
\bibitem [{\citenamefont {Jeffrey}\ \emph {et~al.}(2014)\citenamefont
  {Jeffrey}, \citenamefont {Sank}, \citenamefont {Mutus}, \citenamefont
  {White}, \citenamefont {Kelly}, \citenamefont {Barends}, \citenamefont
  {Chen}, \citenamefont {Chen}, \citenamefont {Chiaro}, \citenamefont
  {Dunsworth} \emph {et~al.}}]{jeffrey2014fast}%
  \BibitemOpen
  \bibfield  {author} {\bibinfo {author} {\bibfnamefont {E.}~\bibnamefont
  {Jeffrey}}, \bibinfo {author} {\bibfnamefont {D.}~\bibnamefont {Sank}},
  \bibinfo {author} {\bibfnamefont {J.}~\bibnamefont {Mutus}}, \bibinfo
  {author} {\bibfnamefont {T.}~\bibnamefont {White}}, \bibinfo {author}
  {\bibfnamefont {J.}~\bibnamefont {Kelly}}, \bibinfo {author} {\bibfnamefont
  {R.}~\bibnamefont {Barends}}, \bibinfo {author} {\bibfnamefont
  {Y.}~\bibnamefont {Chen}}, \bibinfo {author} {\bibfnamefont {Z.}~\bibnamefont
  {Chen}}, \bibinfo {author} {\bibfnamefont {B.}~\bibnamefont {Chiaro}},
  \bibinfo {author} {\bibfnamefont {A.}~\bibnamefont {Dunsworth}},  \emph
  {et~al.},\ }\href@noop {} {\bibfield  {journal} {\bibinfo  {journal} {Phys.
  Rev. Lett.}\ }\textbf {\bibinfo {volume} {112}},\ \bibinfo {pages} {190504}
  (\bibinfo {year} {2014})}\BibitemShut {NoStop}%
\bibitem [{\citenamefont {Bronn}\ \emph {et~al.}(2015)\citenamefont {Bronn},
  \citenamefont {Magesan}, \citenamefont {Masluk}, \citenamefont {Chow},
  \citenamefont {Gambetta},\ and\ \citenamefont {Steffen}}]{bronn2015reducing}%
  \BibitemOpen
  \bibfield  {author} {\bibinfo {author} {\bibfnamefont {N.~T.}\ \bibnamefont
  {Bronn}}, \bibinfo {author} {\bibfnamefont {E.}~\bibnamefont {Magesan}},
  \bibinfo {author} {\bibfnamefont {N.~A.}\ \bibnamefont {Masluk}}, \bibinfo
  {author} {\bibfnamefont {J.~M.}\ \bibnamefont {Chow}}, \bibinfo {author}
  {\bibfnamefont {J.~M.}\ \bibnamefont {Gambetta}}, \ and\ \bibinfo {author}
  {\bibfnamefont {M.}~\bibnamefont {Steffen}},\ }\href@noop {} {\bibfield
  {journal} {\bibinfo  {journal} {IEEE Trans. Appl. Supercond.}\ }\textbf
  {\bibinfo {volume} {25}},\ \bibinfo {pages} {1} (\bibinfo {year}
  {2015})}\BibitemShut {NoStop}%
\bibitem [{\citenamefont {Blais}\ \emph {et~al.}(2004)\citenamefont {Blais},
  \citenamefont {Huang}, \citenamefont {Wallraff}, \citenamefont {Girvin},\
  and\ \citenamefont {Schoelkopf}}]{Blais2004}%
  \BibitemOpen
  \bibfield  {author} {\bibinfo {author} {\bibfnamefont {A.}~\bibnamefont
  {Blais}}, \bibinfo {author} {\bibfnamefont {R.-S.}\ \bibnamefont {Huang}},
  \bibinfo {author} {\bibfnamefont {A.}~\bibnamefont {Wallraff}}, \bibinfo
  {author} {\bibfnamefont {S.~M.}\ \bibnamefont {Girvin}}, \ and\ \bibinfo
  {author} {\bibfnamefont {R.~J.}\ \bibnamefont {Schoelkopf}},\ }\href@noop {}
  {\bibfield  {journal} {\bibinfo  {journal} {Phys. Rev. A}\ }\textbf {\bibinfo
  {volume} {69}},\ \bibinfo {pages} {062320} (\bibinfo {year}
  {2004})}\BibitemShut {NoStop}%
\bibitem [{\citenamefont {Wallraff}\ \emph {et~al.}(2004)\citenamefont
  {Wallraff}, \citenamefont {Schuster}, \citenamefont {Blais}, \citenamefont
  {Frunzio}, \citenamefont {Huang}, \citenamefont {Majer}, \citenamefont
  {Kumar}, \citenamefont {Girvin},\ and\ \citenamefont
  {Schoelkopf}}]{Wallraff2004}%
  \BibitemOpen
  \bibfield  {author} {\bibinfo {author} {\bibfnamefont {A.}~\bibnamefont
  {Wallraff}}, \bibinfo {author} {\bibfnamefont {D.~I.}\ \bibnamefont
  {Schuster}}, \bibinfo {author} {\bibfnamefont {A.}~\bibnamefont {Blais}},
  \bibinfo {author} {\bibfnamefont {L.}~\bibnamefont {Frunzio}}, \bibinfo
  {author} {\bibfnamefont {R.-S.}\ \bibnamefont {Huang}}, \bibinfo {author}
  {\bibfnamefont {J.}~\bibnamefont {Majer}}, \bibinfo {author} {\bibfnamefont
  {S.}~\bibnamefont {Kumar}}, \bibinfo {author} {\bibfnamefont {S.~M.}\
  \bibnamefont {Girvin}}, \ and\ \bibinfo {author} {\bibfnamefont {R.~J.}\
  \bibnamefont {Schoelkopf}},\ }\href@noop {} {\bibfield  {journal} {\bibinfo
  {journal} {Nature}\ }\textbf {\bibinfo {volume} {431}},\ \bibinfo {pages}
  {162} (\bibinfo {year} {2004})}\BibitemShut {NoStop}%
\bibitem [{\citenamefont {Castellanos-Beltran}\ and\ \citenamefont
  {Lehnert}(2007)}]{castellanos2007widely}%
  \BibitemOpen
  \bibfield  {author} {\bibinfo {author} {\bibfnamefont {M.}~\bibnamefont
  {Castellanos-Beltran}}\ and\ \bibinfo {author} {\bibfnamefont
  {K.}~\bibnamefont {Lehnert}},\ }\href@noop {} {\bibfield  {journal} {\bibinfo
   {journal} {Appl. Phys. Lett.}\ }\textbf {\bibinfo {volume} {91}},\ \bibinfo
  {pages} {083509} (\bibinfo {year} {2007})}\BibitemShut {NoStop}%
\bibitem [{\citenamefont {Bergeal}\ \emph {et~al.}(2010)\citenamefont
  {Bergeal}, \citenamefont {Schackert}, \citenamefont {Metcalfe}, \citenamefont
  {Vijay}, \citenamefont {Manucharyan}, \citenamefont {Frunzio}, \citenamefont
  {Prober}, \citenamefont {Schoelkopf}, \citenamefont {Girvin},\ and\
  \citenamefont {Devoret}}]{bergeal2010phase}%
  \BibitemOpen
  \bibfield  {author} {\bibinfo {author} {\bibfnamefont {N.}~\bibnamefont
  {Bergeal}}, \bibinfo {author} {\bibfnamefont {F.}~\bibnamefont {Schackert}},
  \bibinfo {author} {\bibfnamefont {M.}~\bibnamefont {Metcalfe}}, \bibinfo
  {author} {\bibfnamefont {R.}~\bibnamefont {Vijay}}, \bibinfo {author}
  {\bibfnamefont {V.}~\bibnamefont {Manucharyan}}, \bibinfo {author}
  {\bibfnamefont {L.}~\bibnamefont {Frunzio}}, \bibinfo {author} {\bibfnamefont
  {D.}~\bibnamefont {Prober}}, \bibinfo {author} {\bibfnamefont
  {R.}~\bibnamefont {Schoelkopf}}, \bibinfo {author} {\bibfnamefont
  {S.}~\bibnamefont {Girvin}}, \ and\ \bibinfo {author} {\bibfnamefont
  {M.}~\bibnamefont {Devoret}},\ }\href@noop {} {\bibfield  {journal} {\bibinfo
   {journal} {Nature}\ }\textbf {\bibinfo {volume} {465}},\ \bibinfo {pages}
  {64} (\bibinfo {year} {2010})}\BibitemShut {NoStop}%
\bibitem [{\citenamefont {Mutus}\ \emph {et~al.}(2014)\citenamefont {Mutus},
  \citenamefont {White}, \citenamefont {Barends}, \citenamefont {Chen},
  \citenamefont {Chen}, \citenamefont {Chiaro}, \citenamefont {Dunsworth},
  \citenamefont {Jeffrey}, \citenamefont {Kelly}, \citenamefont {Megrant} \emph
  {et~al.}}]{mutus2014strong}%
  \BibitemOpen
  \bibfield  {author} {\bibinfo {author} {\bibfnamefont {J.~Y.}\ \bibnamefont
  {Mutus}}, \bibinfo {author} {\bibfnamefont {T.~C.}\ \bibnamefont {White}},
  \bibinfo {author} {\bibfnamefont {R.}~\bibnamefont {Barends}}, \bibinfo
  {author} {\bibfnamefont {Y.}~\bibnamefont {Chen}}, \bibinfo {author}
  {\bibfnamefont {Z.}~\bibnamefont {Chen}}, \bibinfo {author} {\bibfnamefont
  {B.}~\bibnamefont {Chiaro}}, \bibinfo {author} {\bibfnamefont
  {A.}~\bibnamefont {Dunsworth}}, \bibinfo {author} {\bibfnamefont
  {E.}~\bibnamefont {Jeffrey}}, \bibinfo {author} {\bibfnamefont
  {J.}~\bibnamefont {Kelly}}, \bibinfo {author} {\bibfnamefont
  {A.}~\bibnamefont {Megrant}},  \emph {et~al.},\ }\href@noop {} {\bibfield
  {journal} {\bibinfo  {journal} {Appl. Phys. Lett.}\ }\textbf {\bibinfo
  {volume} {104}},\ \bibinfo {pages} {263513} (\bibinfo {year}
  {2014})}\BibitemShut {NoStop}%
\bibitem [{\citenamefont {Roy}\ \emph {et~al.}(2015)\citenamefont {Roy},
  \citenamefont {Kundu}, \citenamefont {Chand}, \citenamefont {Vadiraj},
  \citenamefont {Ranadive}, \citenamefont {Nehra}, \citenamefont {Patankar},
  \citenamefont {Aumentado}, \citenamefont {Clerk},\ and\ \citenamefont
  {Vijay}}]{roy2015broadband}%
  \BibitemOpen
  \bibfield  {author} {\bibinfo {author} {\bibfnamefont {T.}~\bibnamefont
  {Roy}}, \bibinfo {author} {\bibfnamefont {S.}~\bibnamefont {Kundu}}, \bibinfo
  {author} {\bibfnamefont {M.}~\bibnamefont {Chand}}, \bibinfo {author}
  {\bibfnamefont {A.}~\bibnamefont {Vadiraj}}, \bibinfo {author} {\bibfnamefont
  {A.}~\bibnamefont {Ranadive}}, \bibinfo {author} {\bibfnamefont
  {N.}~\bibnamefont {Nehra}}, \bibinfo {author} {\bibfnamefont {M.~P.}\
  \bibnamefont {Patankar}}, \bibinfo {author} {\bibfnamefont {J.}~\bibnamefont
  {Aumentado}}, \bibinfo {author} {\bibfnamefont {A.}~\bibnamefont {Clerk}}, \
  and\ \bibinfo {author} {\bibfnamefont {R.}~\bibnamefont {Vijay}},\
  }\href@noop {} {\bibfield  {journal} {\bibinfo  {journal} {Appl. Phys.
  Lett.}\ }\textbf {\bibinfo {volume} {107}},\ \bibinfo {pages} {262601}
  (\bibinfo {year} {2015})}\BibitemShut {NoStop}%
\bibitem [{\citenamefont {Bal}\ \emph {et~al.}(2020)\citenamefont {Bal},
  \citenamefont {Long}, \citenamefont {Zhao}, \citenamefont {Wang},
  \citenamefont {Park}, \citenamefont {McRae}, \citenamefont {Zhao},
  \citenamefont {Lake}, \citenamefont {Frolov}, \citenamefont {Pilipenko} \emph
  {et~al.}}]{bal2020overlap}%
  \BibitemOpen
  \bibfield  {author} {\bibinfo {author} {\bibfnamefont {M.}~\bibnamefont
  {Bal}}, \bibinfo {author} {\bibfnamefont {J.}~\bibnamefont {Long}}, \bibinfo
  {author} {\bibfnamefont {R.}~\bibnamefont {Zhao}}, \bibinfo {author}
  {\bibfnamefont {H.}~\bibnamefont {Wang}}, \bibinfo {author} {\bibfnamefont
  {S.}~\bibnamefont {Park}}, \bibinfo {author} {\bibfnamefont {C.~R.~H.}\
  \bibnamefont {McRae}}, \bibinfo {author} {\bibfnamefont {T.}~\bibnamefont
  {Zhao}}, \bibinfo {author} {\bibfnamefont {R.~E.}\ \bibnamefont {Lake}},
  \bibinfo {author} {\bibfnamefont {D.}~\bibnamefont {Frolov}}, \bibinfo
  {author} {\bibfnamefont {R.}~\bibnamefont {Pilipenko}},  \emph {et~al.},\
  }\href@noop {} {\bibfield  {journal} {\bibinfo  {journal} {arXiv preprint
  arXiv:2005.10908}\ } (\bibinfo {year} {2020})}\BibitemShut {NoStop}%
\bibitem [{\citenamefont {Martinis}\ \emph {et~al.}(2005)\citenamefont
  {Martinis}, \citenamefont {Cooper}, \citenamefont {McDermott}, \citenamefont
  {Steffen}, \citenamefont {Ansmann}, \citenamefont {Osborn}, \citenamefont
  {Cicak}, \citenamefont {Oh}, \citenamefont {Pappas}, \citenamefont {Simmonds}
  \emph {et~al.}}]{Martinis2005}%
  \BibitemOpen
  \bibfield  {author} {\bibinfo {author} {\bibfnamefont {J.~M.}\ \bibnamefont
  {Martinis}}, \bibinfo {author} {\bibfnamefont {K.~B.}\ \bibnamefont
  {Cooper}}, \bibinfo {author} {\bibfnamefont {R.}~\bibnamefont {McDermott}},
  \bibinfo {author} {\bibfnamefont {M.}~\bibnamefont {Steffen}}, \bibinfo
  {author} {\bibfnamefont {M.}~\bibnamefont {Ansmann}}, \bibinfo {author}
  {\bibfnamefont {K.}~\bibnamefont {Osborn}}, \bibinfo {author} {\bibfnamefont
  {K.}~\bibnamefont {Cicak}}, \bibinfo {author} {\bibfnamefont
  {S.}~\bibnamefont {Oh}}, \bibinfo {author} {\bibfnamefont {D.~P.}\
  \bibnamefont {Pappas}}, \bibinfo {author} {\bibfnamefont {R.~W.}\
  \bibnamefont {Simmonds}},  \emph {et~al.},\ }\href@noop {} {\bibfield
  {journal} {\bibinfo  {journal} {Phys. Rev. Lett.}\ }\textbf {\bibinfo
  {volume} {95}},\ \bibinfo {pages} {210503} (\bibinfo {year}
  {2005})}\BibitemShut {NoStop}%
\bibitem [{\citenamefont {Gao}(2008)}]{GaoThesis}%
  \BibitemOpen
  \bibfield  {author} {\bibinfo {author} {\bibfnamefont {J.}~\bibnamefont
  {Gao}},\ }\emph {\bibinfo {title} {The physics of superconducting microwave
  resonators}},\ \href@noop {} {Ph.D. thesis},\ \bibinfo  {school} {California
  Institute of Technology} (\bibinfo {year} {2008})\BibitemShut {NoStop}%
\bibitem [{\citenamefont {McRae}\ \emph
  {et~al.}(2020{\natexlab{a}})\citenamefont {McRae}, \citenamefont {Wang},
  \citenamefont {Gao}, \citenamefont {Vissers}, \citenamefont {Brecht},
  \citenamefont {Dunsworth}, \citenamefont {Pappas},\ and\ \citenamefont
  {Mutus}}]{mcrae2020materials}%
  \BibitemOpen
  \bibfield  {author} {\bibinfo {author} {\bibfnamefont {C.~R.~H.}\
  \bibnamefont {McRae}}, \bibinfo {author} {\bibfnamefont {H.}~\bibnamefont
  {Wang}}, \bibinfo {author} {\bibfnamefont {J.}~\bibnamefont {Gao}}, \bibinfo
  {author} {\bibfnamefont {M.~R.}\ \bibnamefont {Vissers}}, \bibinfo {author}
  {\bibfnamefont {T.}~\bibnamefont {Brecht}}, \bibinfo {author} {\bibfnamefont
  {A.}~\bibnamefont {Dunsworth}}, \bibinfo {author} {\bibfnamefont {D.~P.}\
  \bibnamefont {Pappas}}, \ and\ \bibinfo {author} {\bibfnamefont
  {J.}~\bibnamefont {Mutus}},\ }\href@noop {} {\bibfield  {journal} {\bibinfo
  {journal} {Review of Scientific Instruments}\ }\textbf {\bibinfo {volume}
  {91}},\ \bibinfo {pages} {091101} (\bibinfo {year}
  {2020}{\natexlab{a}})}\BibitemShut {NoStop}%
\bibitem [{\citenamefont {Song}\ \emph {et~al.}(2009)\citenamefont {Song},
  \citenamefont {Heitmann}, \citenamefont {DeFeo}, \citenamefont {Yu},
  \citenamefont {McDermott}, \citenamefont {Neeley}, \citenamefont {Martinis},\
  and\ \citenamefont {Plourde}}]{Song2009}%
  \BibitemOpen
  \bibfield  {author} {\bibinfo {author} {\bibfnamefont {C.}~\bibnamefont
  {Song}}, \bibinfo {author} {\bibfnamefont {T.~W.}\ \bibnamefont {Heitmann}},
  \bibinfo {author} {\bibfnamefont {M.~P.}\ \bibnamefont {DeFeo}}, \bibinfo
  {author} {\bibfnamefont {K.}~\bibnamefont {Yu}}, \bibinfo {author}
  {\bibfnamefont {R.}~\bibnamefont {McDermott}}, \bibinfo {author}
  {\bibfnamefont {M.}~\bibnamefont {Neeley}}, \bibinfo {author} {\bibfnamefont
  {J.~M.}\ \bibnamefont {Martinis}}, \ and\ \bibinfo {author} {\bibfnamefont
  {B.~L.}\ \bibnamefont {Plourde}},\ }\href@noop {} {\bibfield  {journal}
  {\bibinfo  {journal} {Phys. Rev. B}\ }\textbf {\bibinfo {volume} {79}},\
  \bibinfo {pages} {174512} (\bibinfo {year} {2009})}\BibitemShut {NoStop}%
\bibitem [{\citenamefont {Chiaro}\ \emph {et~al.}(2016)\citenamefont {Chiaro},
  \citenamefont {Megrant}, \citenamefont {Dunsworth}, \citenamefont {Chen},
  \citenamefont {Barends}, \citenamefont {Campbell}, \citenamefont {Chen},
  \citenamefont {Fowler}, \citenamefont {Hoi}, \citenamefont {Jeffrey} \emph
  {et~al.}}]{Chiaro2016}%
  \BibitemOpen
  \bibfield  {author} {\bibinfo {author} {\bibfnamefont {B.}~\bibnamefont
  {Chiaro}}, \bibinfo {author} {\bibfnamefont {A.}~\bibnamefont {Megrant}},
  \bibinfo {author} {\bibfnamefont {A.}~\bibnamefont {Dunsworth}}, \bibinfo
  {author} {\bibfnamefont {Z.}~\bibnamefont {Chen}}, \bibinfo {author}
  {\bibfnamefont {R.}~\bibnamefont {Barends}}, \bibinfo {author} {\bibfnamefont
  {B.}~\bibnamefont {Campbell}}, \bibinfo {author} {\bibfnamefont
  {Y.}~\bibnamefont {Chen}}, \bibinfo {author} {\bibfnamefont {A.}~\bibnamefont
  {Fowler}}, \bibinfo {author} {\bibfnamefont {I.}~\bibnamefont {Hoi}},
  \bibinfo {author} {\bibfnamefont {E.}~\bibnamefont {Jeffrey}},  \emph
  {et~al.},\ }\href@noop {} {\bibfield  {journal} {\bibinfo  {journal}
  {Supercond. Sci. Technol.}\ }\textbf {\bibinfo {volume} {29}},\ \bibinfo
  {pages} {104006} (\bibinfo {year} {2016})}\BibitemShut {NoStop}%
\bibitem [{\citenamefont {Sage}\ \emph {et~al.}(2011)\citenamefont {Sage},
  \citenamefont {Bolkhovsky}, \citenamefont {Oliver}, \citenamefont {Turek},\
  and\ \citenamefont {Welander}}]{Sage2011}%
  \BibitemOpen
  \bibfield  {author} {\bibinfo {author} {\bibfnamefont {J.~M.}\ \bibnamefont
  {Sage}}, \bibinfo {author} {\bibfnamefont {V.}~\bibnamefont {Bolkhovsky}},
  \bibinfo {author} {\bibfnamefont {W.~D.}\ \bibnamefont {Oliver}}, \bibinfo
  {author} {\bibfnamefont {B.}~\bibnamefont {Turek}}, \ and\ \bibinfo {author}
  {\bibfnamefont {P.~B.}\ \bibnamefont {Welander}},\ }\href@noop {} {\bibfield
  {journal} {\bibinfo  {journal} {J. Appl. Phys.}\ }\textbf {\bibinfo {volume}
  {109}},\ \bibinfo {pages} {063915} (\bibinfo {year} {2011})}\BibitemShut
  {NoStop}%
\bibitem [{\citenamefont {Gao}\ \emph {et~al.}(2008)\citenamefont {Gao},
  \citenamefont {Daal}, \citenamefont {Vayonakis}, \citenamefont {Kumar},
  \citenamefont {Zmuidzinas}, \citenamefont {Sadoulet}, \citenamefont {Mazin},
  \citenamefont {Day},\ and\ \citenamefont {Leduc}}]{Gao2008b}%
  \BibitemOpen
  \bibfield  {author} {\bibinfo {author} {\bibfnamefont {J.}~\bibnamefont
  {Gao}}, \bibinfo {author} {\bibfnamefont {M.}~\bibnamefont {Daal}}, \bibinfo
  {author} {\bibfnamefont {A.}~\bibnamefont {Vayonakis}}, \bibinfo {author}
  {\bibfnamefont {S.}~\bibnamefont {Kumar}}, \bibinfo {author} {\bibfnamefont
  {J.}~\bibnamefont {Zmuidzinas}}, \bibinfo {author} {\bibfnamefont
  {B.}~\bibnamefont {Sadoulet}}, \bibinfo {author} {\bibfnamefont {B.~A.}\
  \bibnamefont {Mazin}}, \bibinfo {author} {\bibfnamefont {P.~K.}\ \bibnamefont
  {Day}}, \ and\ \bibinfo {author} {\bibfnamefont {H.~G.}\ \bibnamefont
  {Leduc}},\ }\href@noop {} {\bibfield  {journal} {\bibinfo  {journal} {Appl.
  Phys. Lett.}\ }\textbf {\bibinfo {volume} {92}},\ \bibinfo {pages} {152505}
  (\bibinfo {year} {2008})}\BibitemShut {NoStop}%
\bibitem [{\citenamefont {McRae}\ \emph
  {et~al.}(2020{\natexlab{b}})\citenamefont {McRae}, \citenamefont {Lake},
  \citenamefont {Long}, \citenamefont {Bal}, \citenamefont {Wu}, \citenamefont
  {Jugdersuren}, \citenamefont {Metcalf}, \citenamefont {Liu},\ and\
  \citenamefont {Pappas}}]{McRae2020}%
  \BibitemOpen
  \bibfield  {author} {\bibinfo {author} {\bibfnamefont {C.~R.~H.}\
  \bibnamefont {McRae}}, \bibinfo {author} {\bibfnamefont {R.~E.}\ \bibnamefont
  {Lake}}, \bibinfo {author} {\bibfnamefont {J.~L.}\ \bibnamefont {Long}},
  \bibinfo {author} {\bibfnamefont {M.}~\bibnamefont {Bal}}, \bibinfo {author}
  {\bibfnamefont {X.}~\bibnamefont {Wu}}, \bibinfo {author} {\bibfnamefont
  {B.}~\bibnamefont {Jugdersuren}}, \bibinfo {author} {\bibfnamefont {T.~H.}\
  \bibnamefont {Metcalf}}, \bibinfo {author} {\bibfnamefont {X.}~\bibnamefont
  {Liu}}, \ and\ \bibinfo {author} {\bibfnamefont {D.~P.}\ \bibnamefont
  {Pappas}},\ }\href {http://arxiv.org/abs/1909.07428} {\bibfield  {journal}
  {\bibinfo  {journal} {Appl. Phys. Lett.}\ }\textbf {\bibinfo {volume}
  {116}},\ \bibinfo {pages} {194003} (\bibinfo {year}
  {2020}{\natexlab{b}})}\BibitemShut {NoStop}%
\bibitem [{\citenamefont {Khalil}\ \emph {et~al.}(2012)\citenamefont {Khalil},
  \citenamefont {Stoutimore}, \citenamefont {Wellstood},\ and\ \citenamefont
  {Osborn}}]{Khalil2012}%
  \BibitemOpen
  \bibfield  {author} {\bibinfo {author} {\bibfnamefont {M.}~\bibnamefont
  {Khalil}}, \bibinfo {author} {\bibfnamefont {M.}~\bibnamefont {Stoutimore}},
  \bibinfo {author} {\bibfnamefont {F.}~\bibnamefont {Wellstood}}, \ and\
  \bibinfo {author} {\bibfnamefont {K.}~\bibnamefont {Osborn}},\ }\href@noop {}
  {\bibfield  {journal} {\bibinfo  {journal} {J. Appl. Phys.}\ }\textbf
  {\bibinfo {volume} {111}},\ \bibinfo {pages} {054510} (\bibinfo {year}
  {2012})}\BibitemShut {NoStop}%
\bibitem [{\citenamefont {Megrant}\ \emph {et~al.}(2012)\citenamefont
  {Megrant}, \citenamefont {Neill}, \citenamefont {Barends}, \citenamefont
  {Chiaro}, \citenamefont {Chen}, \citenamefont {Feigl}, \citenamefont {Kelly},
  \citenamefont {Lucero}, \citenamefont {Mariantoni}, \citenamefont
  {O’Malley} \emph {et~al.}}]{Megrant2012}%
  \BibitemOpen
  \bibfield  {author} {\bibinfo {author} {\bibfnamefont {A.}~\bibnamefont
  {Megrant}}, \bibinfo {author} {\bibfnamefont {C.}~\bibnamefont {Neill}},
  \bibinfo {author} {\bibfnamefont {R.}~\bibnamefont {Barends}}, \bibinfo
  {author} {\bibfnamefont {B.}~\bibnamefont {Chiaro}}, \bibinfo {author}
  {\bibfnamefont {Y.}~\bibnamefont {Chen}}, \bibinfo {author} {\bibfnamefont
  {L.}~\bibnamefont {Feigl}}, \bibinfo {author} {\bibfnamefont
  {J.}~\bibnamefont {Kelly}}, \bibinfo {author} {\bibfnamefont
  {E.}~\bibnamefont {Lucero}}, \bibinfo {author} {\bibfnamefont
  {M.}~\bibnamefont {Mariantoni}}, \bibinfo {author} {\bibfnamefont {P.~J.}\
  \bibnamefont {O’Malley}},  \emph {et~al.},\ }\href@noop {} {\bibfield
  {journal} {\bibinfo  {journal} {Appl. Phys. Lett.}\ }\textbf {\bibinfo
  {volume} {100}},\ \bibinfo {pages} {113510} (\bibinfo {year}
  {2012})}\BibitemShut {NoStop}%
\bibitem [{\citenamefont {Guan}\ \emph {et~al.}(2020)\citenamefont {Guan},
  \citenamefont {Dai}, \citenamefont {He}, \citenamefont {Hu}, \citenamefont
  {Ouyang}, \citenamefont {Wang}, \citenamefont {Wei},\ and\ \citenamefont
  {Gao}}]{guan2020network}%
  \BibitemOpen
  \bibfield  {author} {\bibinfo {author} {\bibfnamefont {H.}~\bibnamefont
  {Guan}}, \bibinfo {author} {\bibfnamefont {M.}~\bibnamefont {Dai}}, \bibinfo
  {author} {\bibfnamefont {Q.}~\bibnamefont {He}}, \bibinfo {author}
  {\bibfnamefont {J.}~\bibnamefont {Hu}}, \bibinfo {author} {\bibfnamefont
  {P.}~\bibnamefont {Ouyang}}, \bibinfo {author} {\bibfnamefont
  {Y.}~\bibnamefont {Wang}}, \bibinfo {author} {\bibfnamefont {L.}~\bibnamefont
  {Wei}}, \ and\ \bibinfo {author} {\bibfnamefont {J.}~\bibnamefont {Gao}},\
  }\href@noop {} {\bibfield  {journal} {\bibinfo  {journal} {Supercond. Sci.
  Technol.}\ }\textbf {\bibinfo {volume} {33}},\ \bibinfo {pages} {075004}
  (\bibinfo {year} {2020})}\BibitemShut {NoStop}%
\bibitem [{\citenamefont {Probst}\ \emph {et~al.}(2015)\citenamefont {Probst},
  \citenamefont {Song}, \citenamefont {Bushev}, \citenamefont {Ustinov},\ and\
  \citenamefont {Weides}}]{Probst2015}%
  \BibitemOpen
  \bibfield  {author} {\bibinfo {author} {\bibfnamefont {S.}~\bibnamefont
  {Probst}}, \bibinfo {author} {\bibfnamefont {F.}~\bibnamefont {Song}},
  \bibinfo {author} {\bibfnamefont {P.}~\bibnamefont {Bushev}}, \bibinfo
  {author} {\bibfnamefont {A.}~\bibnamefont {Ustinov}}, \ and\ \bibinfo
  {author} {\bibfnamefont {M.}~\bibnamefont {Weides}},\ }\href@noop {}
  {\bibfield  {journal} {\bibinfo  {journal} {Rev. Sci. Instrum.}\ }\textbf
  {\bibinfo {volume} {86}},\ \bibinfo {pages} {024706} (\bibinfo {year}
  {2015})}\BibitemShut {NoStop}%
\bibitem [{\citenamefont {Chow}\ \emph {et~al.}(2014)\citenamefont {Chow},
  \citenamefont {Gambetta}, \citenamefont {Magesan}, \citenamefont {Abraham},
  \citenamefont {Cross}, \citenamefont {Johnson}, \citenamefont {Masluk},
  \citenamefont {Ryan}, \citenamefont {Smolin}, \citenamefont {Srinivasan}
  \emph {et~al.}}]{chow2014implementing}%
  \BibitemOpen
  \bibfield  {author} {\bibinfo {author} {\bibfnamefont {J.~M.}\ \bibnamefont
  {Chow}}, \bibinfo {author} {\bibfnamefont {J.~M.}\ \bibnamefont {Gambetta}},
  \bibinfo {author} {\bibfnamefont {E.}~\bibnamefont {Magesan}}, \bibinfo
  {author} {\bibfnamefont {D.~W.}\ \bibnamefont {Abraham}}, \bibinfo {author}
  {\bibfnamefont {A.~W.}\ \bibnamefont {Cross}}, \bibinfo {author}
  {\bibfnamefont {B.}~\bibnamefont {Johnson}}, \bibinfo {author} {\bibfnamefont
  {N.~A.}\ \bibnamefont {Masluk}}, \bibinfo {author} {\bibfnamefont {C.~A.}\
  \bibnamefont {Ryan}}, \bibinfo {author} {\bibfnamefont {J.~A.}\ \bibnamefont
  {Smolin}}, \bibinfo {author} {\bibfnamefont {S.~J.}\ \bibnamefont
  {Srinivasan}},  \emph {et~al.},\ }\href@noop {} {\bibfield  {journal}
  {\bibinfo  {journal} {Nat. Commun.}\ }\textbf {\bibinfo {volume} {5}},\
  \bibinfo {pages} {1} (\bibinfo {year} {2014})}\BibitemShut {NoStop}%
\bibitem [{\citenamefont {Ware}\ \emph {et~al.}(2019)\citenamefont {Ware},
  \citenamefont {Johnson}, \citenamefont {Gambetta}, \citenamefont {Ohki},
  \citenamefont {Chow},\ and\ \citenamefont {Plourde}}]{ware2019cross}%
  \BibitemOpen
  \bibfield  {author} {\bibinfo {author} {\bibfnamefont {M.}~\bibnamefont
  {Ware}}, \bibinfo {author} {\bibfnamefont {B.~R.}\ \bibnamefont {Johnson}},
  \bibinfo {author} {\bibfnamefont {J.~M.}\ \bibnamefont {Gambetta}}, \bibinfo
  {author} {\bibfnamefont {T.~A.}\ \bibnamefont {Ohki}}, \bibinfo {author}
  {\bibfnamefont {J.~M.}\ \bibnamefont {Chow}}, \ and\ \bibinfo {author}
  {\bibfnamefont {B.}~\bibnamefont {Plourde}},\ }\href@noop {} {\bibfield
  {journal} {\bibinfo  {journal} {arXiv preprint arXiv:1905.11480}\ } (\bibinfo
  {year} {2019})}\BibitemShut {NoStop}%
\bibitem [{\citenamefont {Castellanos-Beltran}(2010)}]{manuelthesis}%
  \BibitemOpen
  \bibfield  {author} {\bibinfo {author} {\bibfnamefont {M.~A.}\ \bibnamefont
  {Castellanos-Beltran}},\ }\emph {\bibinfo {title} {Development of a Josephson
  parametric amplifier for the preparation and detection of nonclassical states
  of microwave fields}},\ \href@noop {} {Ph.D. thesis},\ \bibinfo  {school}
  {University of Colorado Boulder} (\bibinfo {year} {2010})\BibitemShut
  {NoStop}%
\bibitem [{\citenamefont {Kudra}\ \emph {et~al.}(2020)\citenamefont {Kudra},
  \citenamefont {Bizn{\'a}rov{\'a}}, \citenamefont {Fadavi~Roudsari},
  \citenamefont {Burnett}, \citenamefont {Niepce}, \citenamefont
  {Gasparinetti}, \citenamefont {Wickman},\ and\ \citenamefont
  {Delsing}}]{kudra2020high}%
  \BibitemOpen
  \bibfield  {author} {\bibinfo {author} {\bibfnamefont {M.}~\bibnamefont
  {Kudra}}, \bibinfo {author} {\bibfnamefont {J.}~\bibnamefont
  {Bizn{\'a}rov{\'a}}}, \bibinfo {author} {\bibfnamefont {A.}~\bibnamefont
  {Fadavi~Roudsari}}, \bibinfo {author} {\bibfnamefont {J.}~\bibnamefont
  {Burnett}}, \bibinfo {author} {\bibfnamefont {D.}~\bibnamefont {Niepce}},
  \bibinfo {author} {\bibfnamefont {S.}~\bibnamefont {Gasparinetti}}, \bibinfo
  {author} {\bibfnamefont {B.}~\bibnamefont {Wickman}}, \ and\ \bibinfo
  {author} {\bibfnamefont {P.}~\bibnamefont {Delsing}},\ }\href@noop {}
  {\bibfield  {journal} {\bibinfo  {journal} {Appl. Phys. Lett.}\ }\textbf
  {\bibinfo {volume} {117}},\ \bibinfo {pages} {070601} (\bibinfo {year}
  {2020})}\BibitemShut {NoStop}%
\bibitem [{\citenamefont {Chakram}\ \emph {et~al.}(2020)\citenamefont
  {Chakram}, \citenamefont {Oriani}, \citenamefont {Naik}, \citenamefont
  {Dixit}, \citenamefont {He}, \citenamefont {Agrawal}, \citenamefont {Kwon},\
  and\ \citenamefont {Schuster}}]{chakram2020seamless}%
  \BibitemOpen
  \bibfield  {author} {\bibinfo {author} {\bibfnamefont {S.}~\bibnamefont
  {Chakram}}, \bibinfo {author} {\bibfnamefont {A.~E.}\ \bibnamefont {Oriani}},
  \bibinfo {author} {\bibfnamefont {R.~K.}\ \bibnamefont {Naik}}, \bibinfo
  {author} {\bibfnamefont {A.~V.}\ \bibnamefont {Dixit}}, \bibinfo {author}
  {\bibfnamefont {K.}~\bibnamefont {He}}, \bibinfo {author} {\bibfnamefont
  {A.}~\bibnamefont {Agrawal}}, \bibinfo {author} {\bibfnamefont
  {H.}~\bibnamefont {Kwon}}, \ and\ \bibinfo {author} {\bibfnamefont {D.~I.}\
  \bibnamefont {Schuster}},\ }\href@noop {} {\bibfield  {journal} {\bibinfo
  {journal} {arXiv preprint arXiv:2010.16382}\ } (\bibinfo {year}
  {2020})}\BibitemShut {NoStop}%
\bibitem [{\citenamefont {Jun}\ \emph {et~al.}(2004)\citenamefont {Jun},
  \citenamefont {Hwang}, \citenamefont {Jeong},\ and\ \citenamefont
  {Ahn}}]{jun2004microwave}%
  \BibitemOpen
  \bibfield  {author} {\bibinfo {author} {\bibfnamefont {M.}~\bibnamefont
  {Jun}}, \bibinfo {author} {\bibfnamefont {S.}~\bibnamefont {Hwang}}, \bibinfo
  {author} {\bibfnamefont {D.}~\bibnamefont {Jeong}}, \ and\ \bibinfo {author}
  {\bibfnamefont {D.}~\bibnamefont {Ahn}},\ }\href@noop {} {\bibfield
  {journal} {\bibinfo  {journal} {Rev. Sci. Instrum.}\ }\textbf {\bibinfo
  {volume} {75}},\ \bibinfo {pages} {2455} (\bibinfo {year}
  {2004})}\BibitemShut {NoStop}%
\bibitem [{\citenamefont {Slichter}, \citenamefont {Naaman},\ and\
  \citenamefont {Siddiqi}(2009)}]{slichter2009millikelvin}%
  \BibitemOpen
  \bibfield  {author} {\bibinfo {author} {\bibfnamefont {D.}~\bibnamefont
  {Slichter}}, \bibinfo {author} {\bibfnamefont {O.}~\bibnamefont {Naaman}}, \
  and\ \bibinfo {author} {\bibfnamefont {I.}~\bibnamefont {Siddiqi}},\
  }\href@noop {} {\bibfield  {journal} {\bibinfo  {journal} {Appl. Phys.
  Lett.}\ }\textbf {\bibinfo {volume} {94}},\ \bibinfo {pages} {192508}
  (\bibinfo {year} {2009})}\BibitemShut {NoStop}%
\bibitem [{\citenamefont {Ranzani}\ \emph {et~al.}(2013)\citenamefont
  {Ranzani}, \citenamefont {Spietz}, \citenamefont {Popovic},\ and\
  \citenamefont {Aumentado}}]{ranzani2013two}%
  \BibitemOpen
  \bibfield  {author} {\bibinfo {author} {\bibfnamefont {L.}~\bibnamefont
  {Ranzani}}, \bibinfo {author} {\bibfnamefont {L.}~\bibnamefont {Spietz}},
  \bibinfo {author} {\bibfnamefont {Z.}~\bibnamefont {Popovic}}, \ and\
  \bibinfo {author} {\bibfnamefont {J.}~\bibnamefont {Aumentado}},\ }\href@noop
  {} {\bibfield  {journal} {\bibinfo  {journal} {Rev. Sci. Instrum.}\ }\textbf
  {\bibinfo {volume} {84}},\ \bibinfo {pages} {034704} (\bibinfo {year}
  {2013})}\BibitemShut {NoStop}%
\bibitem [{\citenamefont {Yeh}\ and\ \citenamefont
  {Anlage}(2013)}]{yeh2013situ}%
  \BibitemOpen
  \bibfield  {author} {\bibinfo {author} {\bibfnamefont {J.-H.}\ \bibnamefont
  {Yeh}}\ and\ \bibinfo {author} {\bibfnamefont {S.~M.}\ \bibnamefont
  {Anlage}},\ }\href@noop {} {\bibfield  {journal} {\bibinfo  {journal} {Rev.
  Sci. Instrum.}\ }\textbf {\bibinfo {volume} {84}},\ \bibinfo {pages} {034706}
  (\bibinfo {year} {2013})}\BibitemShut {NoStop}%
\bibitem [{app(2019{\natexlab{a}})}]{app-note2}%
  \BibitemOpen
  \href@noop {} {\enquote {\bibinfo {title} {Network analyzer architectures},}\
  }\bibinfo {howpublished} {Available at
  \url{https://www.keysight.com/us/en/assets/3119-1082/application-notes/5965-7708.pdf}}
  (\bibinfo {year} {2019}{\natexlab{a}})\BibitemShut {NoStop}%
\bibitem [{\citenamefont {Rytting}(1980)}]{rytting1980analysis}%
  \BibitemOpen
  \bibfield  {author} {\bibinfo {author} {\bibfnamefont {D.}~\bibnamefont
  {Rytting}},\ }in\ \href@noop {} {\emph {\bibinfo {booktitle} {RF and
  Microwave Symp. Exhibition}}}\ (\bibinfo {year} {1980})\BibitemShut {NoStop}%
\bibitem [{\citenamefont {Garelli}\ and\ \citenamefont
  {Ferrero}(2012)}]{garelli2012unified}%
  \BibitemOpen
  \bibfield  {author} {\bibinfo {author} {\bibfnamefont {M.}~\bibnamefont
  {Garelli}}\ and\ \bibinfo {author} {\bibfnamefont {A.}~\bibnamefont
  {Ferrero}},\ }\href@noop {} {\bibfield  {journal} {\bibinfo  {journal} {IEEE
  transactions on microwave theory and techniques}\ }\textbf {\bibinfo {volume}
  {60}},\ \bibinfo {pages} {3844} (\bibinfo {year} {2012})}\BibitemShut
  {NoStop}%
\bibitem [{app(2019{\natexlab{b}})}]{app-note1}%
  \BibitemOpen
  \href@noop {} {\enquote {\bibinfo {title} {Specifying calibration standards
  and kits for keysight vector network analyzers},}\ }\bibinfo {howpublished}
  {Available at
  \url{https://www.keysight.com/us/en/assets/7018-01375/application-notes/5989-4840.pdf}}
  (\bibinfo {year} {2019}{\natexlab{b}})\BibitemShut {NoStop}%
\bibitem [{\citenamefont {Bianco}\ \emph {et~al.}(1978)\citenamefont {Bianco},
  \citenamefont {Corana}, \citenamefont {Ridella},\ and\ \citenamefont
  {Simicich}}]{bianco1978evaluation}%
  \BibitemOpen
  \bibfield  {author} {\bibinfo {author} {\bibfnamefont {B.}~\bibnamefont
  {Bianco}}, \bibinfo {author} {\bibfnamefont {A.}~\bibnamefont {Corana}},
  \bibinfo {author} {\bibfnamefont {S.}~\bibnamefont {Ridella}}, \ and\
  \bibinfo {author} {\bibfnamefont {C.}~\bibnamefont {Simicich}},\ }\href@noop
  {} {\bibfield  {journal} {\bibinfo  {journal} {IEEE T. Instrum. Meas.}\
  }\textbf {\bibinfo {volume} {27}},\ \bibinfo {pages} {354} (\bibinfo {year}
  {1978})}\BibitemShut {NoStop}%
\bibitem [{\citenamefont {Blackham}\ and\ \citenamefont
  {Wong}(2005)}]{blackham2005latest}%
  \BibitemOpen
  \bibfield  {author} {\bibinfo {author} {\bibfnamefont {D.}~\bibnamefont
  {Blackham}}\ and\ \bibinfo {author} {\bibfnamefont {K.}~\bibnamefont
  {Wong}},\ }\href@noop {} {\bibfield  {journal} {\bibinfo  {journal}
  {Microwave Journal}\ }\textbf {\bibinfo {volume} {48}} (\bibinfo {year}
  {2005})}\BibitemShut {NoStop}%
\bibitem [{Note1()}]{Note1}%
  \BibitemOpen
  \bibinfo {note}
  {Https://github.com/Boulder-Cryogenic-Quantum-Testbed/measurement/}\BibitemShut
  {NoStop}%
\bibitem [{\citenamefont {Bruno}\ \emph {et~al.}(2015)\citenamefont {Bruno},
  \citenamefont {De~Lange}, \citenamefont {Asaad}, \citenamefont {Van
  Der~Enden}, \citenamefont {Langford},\ and\ \citenamefont
  {DiCarlo}}]{Bruno2015}%
  \BibitemOpen
  \bibfield  {author} {\bibinfo {author} {\bibfnamefont {A.}~\bibnamefont
  {Bruno}}, \bibinfo {author} {\bibfnamefont {G.}~\bibnamefont {De~Lange}},
  \bibinfo {author} {\bibfnamefont {S.}~\bibnamefont {Asaad}}, \bibinfo
  {author} {\bibfnamefont {K.}~\bibnamefont {Van Der~Enden}}, \bibinfo {author}
  {\bibfnamefont {N.}~\bibnamefont {Langford}}, \ and\ \bibinfo {author}
  {\bibfnamefont {L.}~\bibnamefont {DiCarlo}},\ }\href@noop {} {\bibfield
  {journal} {\bibinfo  {journal} {Appl. Phys. Lett.}\ }\textbf {\bibinfo
  {volume} {106}},\ \bibinfo {pages} {182601} (\bibinfo {year}
  {2015})}\BibitemShut {NoStop}%
\bibitem [{\citenamefont {Erickson}\ \emph {et~al.}(2014)\citenamefont
  {Erickson}, \citenamefont {Vissers}, \citenamefont {Sandberg}, \citenamefont
  {Jefferts},\ and\ \citenamefont {Pappas}}]{erickson2014frequency}%
  \BibitemOpen
  \bibfield  {author} {\bibinfo {author} {\bibfnamefont {R.~P.}\ \bibnamefont
  {Erickson}}, \bibinfo {author} {\bibfnamefont {M.~R.}\ \bibnamefont
  {Vissers}}, \bibinfo {author} {\bibfnamefont {M.}~\bibnamefont {Sandberg}},
  \bibinfo {author} {\bibfnamefont {S.~R.}\ \bibnamefont {Jefferts}}, \ and\
  \bibinfo {author} {\bibfnamefont {D.~P.}\ \bibnamefont {Pappas}},\
  }\href@noop {} {\bibfield  {journal} {\bibinfo  {journal} {Phys. Rev. Lett.}\
  }\textbf {\bibinfo {volume} {113}},\ \bibinfo {pages} {187002} (\bibinfo
  {year} {2014})}\BibitemShut {NoStop}%
\bibitem [{\citenamefont {Richardson}\ \emph {et~al.}(2016)\citenamefont
  {Richardson}, \citenamefont {Siwak}, \citenamefont {Hackley}, \citenamefont
  {Keane}, \citenamefont {Robinson}, \citenamefont {Arey}, \citenamefont
  {Arslan},\ and\ \citenamefont {Palmer}}]{Richardson2016}%
  \BibitemOpen
  \bibfield  {author} {\bibinfo {author} {\bibfnamefont {C.}~\bibnamefont
  {Richardson}}, \bibinfo {author} {\bibfnamefont {N.}~\bibnamefont {Siwak}},
  \bibinfo {author} {\bibfnamefont {J.}~\bibnamefont {Hackley}}, \bibinfo
  {author} {\bibfnamefont {Z.}~\bibnamefont {Keane}}, \bibinfo {author}
  {\bibfnamefont {J.}~\bibnamefont {Robinson}}, \bibinfo {author}
  {\bibfnamefont {B.}~\bibnamefont {Arey}}, \bibinfo {author} {\bibfnamefont
  {I.}~\bibnamefont {Arslan}}, \ and\ \bibinfo {author} {\bibfnamefont
  {B.}~\bibnamefont {Palmer}},\ }\href@noop {} {\bibfield  {journal} {\bibinfo
  {journal} {Supercond. Sci. Technol.}\ }\textbf {\bibinfo {volume} {29}},\
  \bibinfo {pages} {064003} (\bibinfo {year} {2016})}\BibitemShut {NoStop}%
\bibitem [{\citenamefont {Kalacheva}\ \emph {et~al.}(2020)\citenamefont
  {Kalacheva}, \citenamefont {Fedorov}, \citenamefont {Kulakova}, \citenamefont
  {Zotova}, \citenamefont {Korostylev}, \citenamefont {Khrapach}, \citenamefont
  {Ustinov},\ and\ \citenamefont {Astafiev}}]{kalacheva2020improving}%
  \BibitemOpen
  \bibfield  {author} {\bibinfo {author} {\bibfnamefont {D.}~\bibnamefont
  {Kalacheva}}, \bibinfo {author} {\bibfnamefont {G.}~\bibnamefont {Fedorov}},
  \bibinfo {author} {\bibfnamefont {A.}~\bibnamefont {Kulakova}}, \bibinfo
  {author} {\bibfnamefont {J.}~\bibnamefont {Zotova}}, \bibinfo {author}
  {\bibfnamefont {E.}~\bibnamefont {Korostylev}}, \bibinfo {author}
  {\bibfnamefont {I.}~\bibnamefont {Khrapach}}, \bibinfo {author}
  {\bibfnamefont {A.}~\bibnamefont {Ustinov}}, \ and\ \bibinfo {author}
  {\bibfnamefont {O.}~\bibnamefont {Astafiev}},\ }in\ \href@noop {} {\emph
  {\bibinfo {booktitle} {AIP Conference Proceedings}}},\ Vol.\ \bibinfo
  {volume} {2241}\ (\bibinfo {organization} {AIP Publishing LLC},\ \bibinfo
  {year} {2020})\ p.\ \bibinfo {pages} {020018}\BibitemShut {NoStop}%
\bibitem [{\citenamefont {Wang}\ \emph {et~al.}(2019)\citenamefont {Wang},
  \citenamefont {Zhuravel}, \citenamefont {Indrajeet}, \citenamefont
  {Taketani}, \citenamefont {Hutchings}, \citenamefont {Hao}, \citenamefont
  {Rouxinol}, \citenamefont {Wilhelm}, \citenamefont {LaHaye}, \citenamefont
  {Ustinov} \emph {et~al.}}]{wang2019mode}%
  \BibitemOpen
  \bibfield  {author} {\bibinfo {author} {\bibfnamefont {H.}~\bibnamefont
  {Wang}}, \bibinfo {author} {\bibfnamefont {A.}~\bibnamefont {Zhuravel}},
  \bibinfo {author} {\bibfnamefont {S.}~\bibnamefont {Indrajeet}}, \bibinfo
  {author} {\bibfnamefont {B.~G.}\ \bibnamefont {Taketani}}, \bibinfo {author}
  {\bibfnamefont {M.}~\bibnamefont {Hutchings}}, \bibinfo {author}
  {\bibfnamefont {Y.}~\bibnamefont {Hao}}, \bibinfo {author} {\bibfnamefont
  {F.}~\bibnamefont {Rouxinol}}, \bibinfo {author} {\bibfnamefont
  {F.}~\bibnamefont {Wilhelm}}, \bibinfo {author} {\bibfnamefont
  {M.}~\bibnamefont {LaHaye}}, \bibinfo {author} {\bibfnamefont
  {A.}~\bibnamefont {Ustinov}},  \emph {et~al.},\ }\href@noop {} {\bibfield
  {journal} {\bibinfo  {journal} {Phys. Rev. Appl.}\ }\textbf {\bibinfo
  {volume} {11}},\ \bibinfo {pages} {054062} (\bibinfo {year}
  {2019})}\BibitemShut {NoStop}%
\bibitem [{\citenamefont {Pappas}\ \emph {et~al.}(2011)\citenamefont {Pappas},
  \citenamefont {Vissers}, \citenamefont {Wisbey}, \citenamefont {Kline},\ and\
  \citenamefont {Gao}}]{Pappas2011}%
  \BibitemOpen
  \bibfield  {author} {\bibinfo {author} {\bibfnamefont {D.~P.}\ \bibnamefont
  {Pappas}}, \bibinfo {author} {\bibfnamefont {M.~R.}\ \bibnamefont {Vissers}},
  \bibinfo {author} {\bibfnamefont {D.~S.}\ \bibnamefont {Wisbey}}, \bibinfo
  {author} {\bibfnamefont {J.~S.}\ \bibnamefont {Kline}}, \ and\ \bibinfo
  {author} {\bibfnamefont {J.}~\bibnamefont {Gao}},\ }\href@noop {} {\bibfield
  {journal} {\bibinfo  {journal} {IEEE Trans. Appl. Supercond.}\ }\textbf
  {\bibinfo {volume} {21}},\ \bibinfo {pages} {871} (\bibinfo {year}
  {2011})}\BibitemShut {NoStop}%
\bibitem [{\citenamefont {Burnett}\ \emph {et~al.}(2018)\citenamefont
  {Burnett}, \citenamefont {Bengtsson}, \citenamefont {Niepce},\ and\
  \citenamefont {Bylander}}]{Burnett2018}%
  \BibitemOpen
  \bibfield  {author} {\bibinfo {author} {\bibfnamefont {J.}~\bibnamefont
  {Burnett}}, \bibinfo {author} {\bibfnamefont {A.}~\bibnamefont {Bengtsson}},
  \bibinfo {author} {\bibfnamefont {D.}~\bibnamefont {Niepce}}, \ and\ \bibinfo
  {author} {\bibfnamefont {J.}~\bibnamefont {Bylander}},\ }\href@noop {}
  {\bibfield  {journal} {\bibinfo  {journal} {J. Phys. Conf. Ser.}\ }\textbf
  {\bibinfo {volume} {969}},\ \bibinfo {pages} {012131} (\bibinfo {year}
  {2018})}\BibitemShut {NoStop}%
\end{thebibliography}%

\end{document}